\documentclass[10pt,journal]{IEEEtran}
\usepackage{graphicx}
\usepackage{setspace}

\usepackage{color}
\usepackage{lineno}
\usepackage{subfigure}

\usepackage{url}
\ifCLASSOPTIONcompsoc
  \usepackage[nocompress]{cite}
\else
  \usepackage{cite}
\fi
\usepackage{amssymb}
\usepackage{algorithmic}
\usepackage{algorithm}
\usepackage[cmex10]{amsmath}
\usepackage[para]{threeparttable}
\usepackage{array}

\usepackage{multirow}
\usepackage{bigdelim}
\usepackage{bigstrut}

\begin{document}
\title{Approaching Single-Hop Performance in Multi-Hop Networks: End-To-End Known-Interference Cancellation (E2E-KIC)}

\author{Fanzhao~Wang,~\IEEEmembership{Student~Member,~IEEE,}
        Lei~Guo\footnote{Lei Guo was the corresponding author for this article.},~\IEEEmembership{Member,~IEEE,}
        Shiqiang~Wang,~\IEEEmembership{Student~Member,~IEEE,}
        Qingyang~Song,~\IEEEmembership{Senior~Member,~IEEE,}
        and~Abbas~Jamalipour,~\IEEEmembership{Fellow,~IEEE}% <-this % stops a space
\IEEEcompsocitemizethanks{\IEEEcompsocthanksitem Copyright (c) 2015 IEEE. Personal use of this material is permitted. However, permission to use this material for any other purposes must be obtained from the IEEE by sending a request to pubs-permissions@ieee.org.
\IEEEcompsocthanksitem F. Wang, L. Guo, and Q. Song are with the School of Information Science and Engineering, Northeastern University, Shenyang 110819, P. R. China. E-mail: fanzhaowang@gmail.com, guolei@ise.neu.edu.cn,  songqingyang@ise.neu.edu.cn
\IEEEcompsocthanksitem S. Wang is with the Department of Electrical and Electronic Engineering, Imperial College London, SW7 2AZ, United Kingdom. E-mail: shiqiang.wang11@imperial.ac.uk
\IEEEcompsocthanksitem A. Jamalipour is with the School of Electrical and Information Engineering, University of Sydney, NSW, 2006, Australia. E-mail: a.jamalipour@ieee.org}
%\vspace{-0.6in}
}% <-this % stops an unwanted space
%\thanks{}

\maketitle
\thispagestyle{plain}
\pagestyle{plain}

%\IEEEtitleabstractindextext{%
\begin{abstract}
To improve the efficiency of wireless data communications, new physical-layer transmission methods based on known-interference cancellation (KIC) have been developed. These methods share the common idea that the interference can be cancelled when the content of it is known. Existing work on KIC mainly focuses on single-hop or two-hop networks, with physical-layer network coding (PNC) and full-duplex (FD) communications as typical examples. This paper extends the idea of KIC to general multi-hop networks, and proposes an end-to-end KIC (E2E-KIC) transmission method together with its MAC design. With E2E-KIC, multiple nodes in a flow passing through a few nodes in an arbitrary topology can simultaneously transmit and receive on the same channel. We first present a theoretical analysis on the effectiveness of E2E-KIC in an idealized case. Then, to support E2E-KIC in multi-hop networks with arbitrary topology, we propose an E2E-KIC-supported MAC protocol (E2E-KIC MAC), which is based on an extension of the Request-to-Send/Clear-to-Send (RTS/CTS) mechanism in the IEEE 802.11 MAC. We also analytically analyze the performance of the proposed E2E-KIC MAC in the presence of hidden terminals. Simulation results illustrate that the proposed E2E-KIC MAC protocol can improve the network throughput and reduce the end-to-end delay.
\end{abstract}

%}

%\IEEEdisplaynontitleabstractindextext

% \fi
%\IEEEpeerreviewmaketitle
%\linenumbers

%\begin{spacing}{1.76}
%\IEEEraisesectionheading{
\section{Introduction}\label{sec:introduction}
%}
%\IEEEPARstart{W}{ireless}
Wireless multi-hop networks, such as wireless ad hoc and sensor networks, are attracting increasing attention because they can increase network coverage, reduce power consumption, and can be easily deployed at low cost \cite{MultiHop}. Traditionally, wireless multi-hop networks use the store-and-forward method (referred to as plain routing, PR, in this paper) for packet relaying, which generally has low throughput, long delay, and high packet loss, particularly when the number of hops is large \cite{Problems2}. Recently, known-interference cancellation (KIC) based technologies emerged as a promising method to improve the performance of wireless multi-hop networks \cite{OCSMA,KIC2,BKIC,refHotTopic,SMAC,fullDuplex}. When receiving a signal with superposed interferences, a node knows the contents of these interferences if they were either received, overheard, or generated by the node. These known interferences can be effectively cancelled by some newly emerged physical-layer techniques \cite{refHotTopic,SMAC,fullDuplex}. It turns out that more concurrent transmissions are allowed with KIC.

\begin{figure}[!t]
\centering \includegraphics[width=0.49\textwidth]{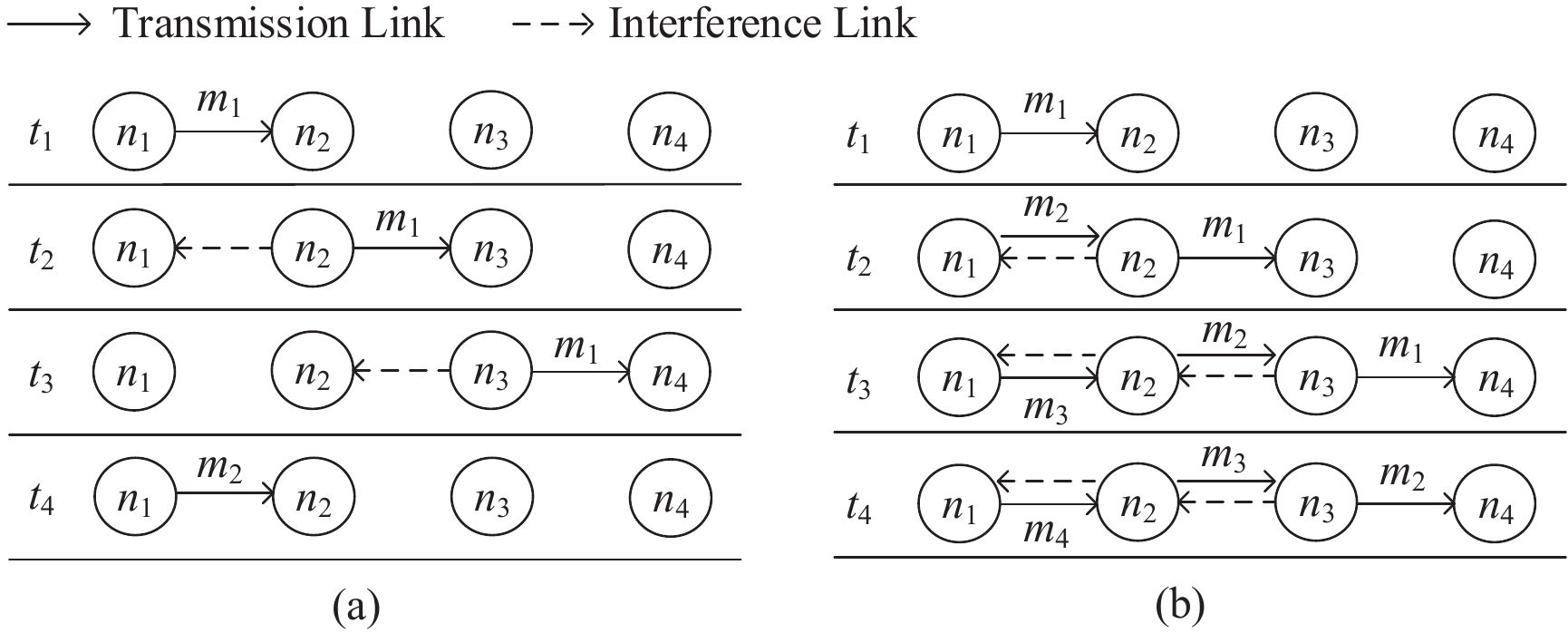}
\caption{Flow through four neighboring nodes transmitted by PR and E2E-KIC: (a) PR, (b) E2E-KIC. Variables $t$, $n$ and $m$ respectively denote timeslots, nodes and packets. One data flow from $n_{1}$ to $n_{4}$ is configured.}
\label{InvolvedRelayings}
\vspace*{-1pt}
\end{figure}

Existing work on KIC mainly focused on limited hops (typically, one or two hops \cite{refHotTopic,SMAC,fullDuplex}). In this paper, we study how to apply KIC in general \emph{multi-hop} cases and propose a new type of KIC, i.e., \emph{end-to-end KIC (E2E-KIC)}. \figurename~\ref{InvolvedRelayings} illustrates the basic idea of E2E-KIC with comparison to PR, where packets are sent from node $n_{1}$ to $n_{4}$. The example flow $n_{1} \to n_{4}$ can be regarded as one particular flow passing through a few nodes in an arbitrary topology. The PR method is shown in Fig. 1(a), where nodes operate in half-duplex mode and the packets are forwarded one-by-one. Fig. 1(b) shows the E2E-KIC method. We take the receptions and transmissions of $n_{2}$ in timeslot $t_{3}$ as an example. In this timeslot, node $n_{2}$ receives three signals from $n_{1}$, $n_{3}$ and itself (i.e., a superposition of the signals $x_{3}$, $x_{2}$, and $x_{1}$ (respectively from $n_{1}$, $n_{2}$, and $n_{3}$ that carry packets $m_{3}$, $m_{2}$, and $m_{1}$)). To successfully exact its intended signal $x_{3}$, node $n_{2}$ should cancel the interfering signals $x_{1}$ and $x_{2}$. Fortunately, $n_{2}$ knows these two interfering signals from the previous transmissions in timeslots $t_{1}$ and $t_{2}$, so they can be cancelled using KIC technologies. We can also see from Fig. 1(b) that the source node transmits one packet in each timeslot, thus the throughput of E2E-KIC can be potentially \emph{the same as single-hop communication}. We will discuss these issues in more details in Section \ref{secMotivation}. E2E-KIC can be used in various practical scenarios, e.g. vehicle-to-vehicle communications (a typical example scenario for wireless multi-hop networks) \cite{V2V}.

Because multiple nodes need to be coordinated to access the channel, it is necessary to develop a \emph{distributed MAC protocol} for E2E-KIC. We face the following challenges in the design of the E2E-KIC-supported MAC protocol.
\begin{enumerate}
\item Ensure that interferences are known. Nodes involved in the E2E-KIC process should know at least the contents of interferences from their neighbors to successfully decode their wanted packets.
\item Control the number of involved nodes. Theoretically, E2E-KIC can support any number of nodes. However, more nodes means higher collision probability and coordination complexity. Therefore, it is necessary to control the number of nodes to maintain efficient E2E-KIC with low collision probability and coordination complexity.
\item Avoid unnecessary contentions. With E2E-KIC, nodes in a multi-hop flow can simultaneously transmit. Thus, some nodes do not need to contend for the wireless channel.
\item Reduce control overhead. It is necessary to inform nodes to participate in an E2E-KIC process with low control overhead, so that the benefit of E2E-KIC can be maintained.
\end{enumerate}

To tackle these challenges, we propose an E2E-KIC MAC protocol, which is based on an extension of the Request-to-Send/Clear-to-Send (RTS/CTS) mechanism in the IEEE 802.11 MAC protocol \cite{M80211}. Nodes randomly access the channel according to a new access-control algorithm that is designed to reduce unnecessary contentions. Through the exchange of RTS and CTS frames, nodes in a multi-hop flow are coordinated to participate in an E2E-KIC process. Because interferences can be effectively cancelled, these nodes can simultaneously transmit their data frames in the E2E-KIC process. The core idea of E2E-KIC MAC is to encourage known interference which can be used for E2E-KIC and avoid unknown interference that may result in packet loss. We also analyze the performance of E2E-KIC MAC with the consideration of hidden terminals, multiple concurrent transmissions, and multi-hop flows. Due to space limitation, an extended version of this paper can be found in our technical report \cite{OurTechRep}, which includes more details on the MAC design and performance analysis.

The remainder of this paper is organized as follows. Section \ref{secRelatedWork} summarizes the related work. Section \ref{secMotivation} analyzes the signal transmission and transmission efficiency of E2E-KIC. Section \ref{secOverviewMAC} overviews the E2E-KIC MAC protocol. Section \ref{sectMAC} presents its details. Section \ref{secThroughputAna} analyzes the performance of E2E-KIC MAC. The simulation results are given in Section \ref{sectSim}. Section \ref{sectCon} draws conclusions and outlines the future work.
\begin{figure}[!t]
\centering \includegraphics[width=0.49\textwidth]{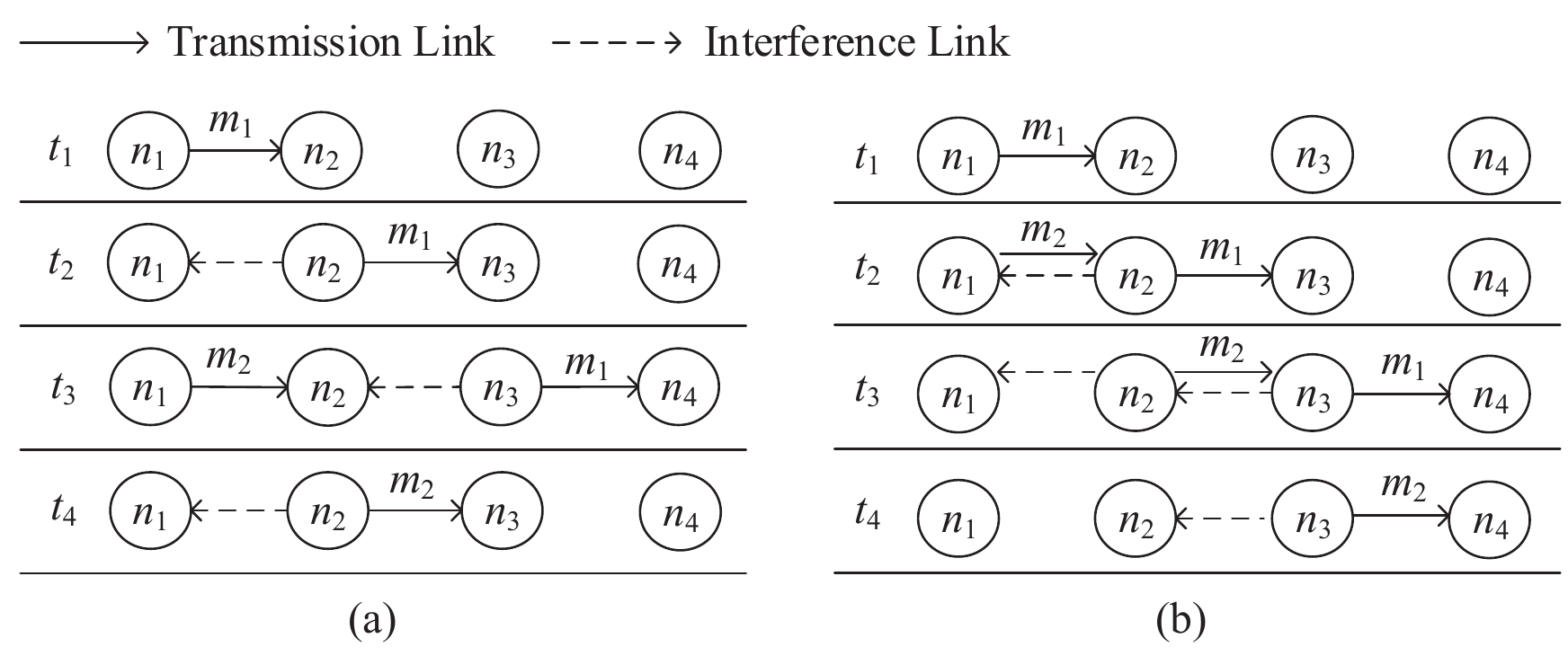}
\caption{Flow through four neighboring nodes transmitted by existing KIC technologies: (a) PNC, (b) FD. Variables $t$, $n$ and $m$ respectively denote timeslots, nodes and packets. One data flow from $n_{1}$ to $n_{4}$ is configured.}
\label{PNCandFD}
\vspace*{-1pt}
\end{figure}
\section{Related Work}\label{secRelatedWork}
Cooperative communications can increase the opportunity of simultaneous transmissions in wireless networks \cite{OCSMA,KIC2,BKIC,DongDelay,MinimizingCostWang,NCBroadcastingWang}. The idea of KIC is an emerging cooperative communication method \cite{OCSMA,KIC2,BKIC}. Physical-layer network coding (PNC) and full-duplex (FD) communications are two typical examples of KIC \cite{refHotTopic,SMAC,fullDuplex}. Example scenarios of PNC and FD are shown in Fig. \ref{PNCandFD}, where node $n_{1}$ intends to send packets to node $n_{4}$. \figurename~\ref{PNCandFD}(a) depicts the idea of PNC with one data flow as proposed in \cite{OCSMA}. A transmission from $n_{1}$ to $n_{2}$ can coexist with the transmission from $n_{3}$ to $n_{4}$ in timeslot $t_3$. PNC can also be used with multiple data flows \cite{refHotTopic,SMAC,OptimalFadingPNC}, where two nodes want to exchange packets through relay nodes. FD is shown in \figurename~\ref{PNCandFD}(b), where nodes can receive and transmit simultaneously by using the self-interference cancellation technology~\cite{fullDuplex}. From Fig. \ref{PNCandFD}, we can see that PNC and FD are different from the proposed E2E-KIC, because they focus on KIC within one or two hops. More application examples of KIC can be found in \cite{KIC2}, which also focus on KIC with one or two hops.

To schedule simultaneous transmissions of wireless nodes, it is necessary to design KIC-supported MAC protocols. Existing KIC-supported MAC protocols include two groups: 1) PNC-supported MAC protocols; 2) FD-supported MAC protocols.

Regarding PNC-supported MAC protocols, most work only focused on two-hop scenarios \cite{refMAC18,refMAC20}, and it is not straightforward to extend these mechanisms to general multi-hop networks. A distributed MAC protocol supporting PNC with one data flow was designed in \cite{OCSMA}, which can work in general multi-hop networks. Some MAC protocols have also been developed to support the case where PNC with multiple data flows exist \cite{MAC,GC_1}. These protocols consider queuing issues and interactions between nodes in multi-hop networks. They can effectively support PNC with multiple data flows.

Regarding FD-supported MAC protocols, the main purpose of works in \cite{Achieving, Rethinking, PracticalFD, MuraFD, UAVFD, sundaresan2014full} is to verify the feasibility of FD radio in practice, with a little modification on the IEEE 802.11 MAC protocol to enable the FD function. These MAC protocols are not capable of working in large multi-hop networks due to the existence of unnecessary channel contentions. Considering this problem, an FD-supported MAC protocol was proposed in \cite{Limits}, which has some channel-sharing mechanisms to reduce collisions and increase FD opportunities. An efficient and fair FD (EF-FD) MAC protocol was proposed in \cite{EFFD}. With a distributed algorithm, EF-FD can reduce unnecessary channel contentions and increase the success probability of FD.

The theoretical performance analysis for MAC protocols operating in wireless multi-hop networks is a challenging topic, especially for networks with hidden terminals. Existing work mainly focuses on special scenarios, such as two-hop topology without hidden terminals \cite{Bianchi1}, \cite[Appendix]{ANCMAC}, two-hop topology with hidden terminals \cite{jang2012ieee}, line topology \cite{ng2007throughput}, etc. Little work focuses on general multi-hop topology with hidden terminals \cite{tsertou2005new,ThroughputTingChao}. Compared with existing work in \cite{tsertou2005new,ThroughputTingChao}, our analysis of E2E-KIC MAC in a general multi-hop topology with hidden terminals is more challenging, because we need to consider the concurrent transmission of multiple nodes that belong to the same multi-hop flow (due to KIC operation), whereas one-hop flows are assumed in \cite{tsertou2005new,ThroughputTingChao} without considering concurrent transmissions that are caused by KIC operation.
\section{End-To-End Known-Interference Cancellation}\label{secMotivation}
In this section, we first analyze the signal transmissions with E2E-KIC to show its feasibility in physical layer. Then, the transmission efficiencies of PR, PNC, FD, and E2E-KIC are discussed. In the following analysis, we consider a typical scenario for KIC, i.e., a flow passing through $N~(N\geq 4)$ nodes in an ad hoc network, where the first node $n_{1}$ wants to send packets to the last node $n_{N}$.

\subsection{Signal Analysis}\label{SignalAna}

\begin{figure}[!t]
\centering \includegraphics[width=0.49\textwidth]{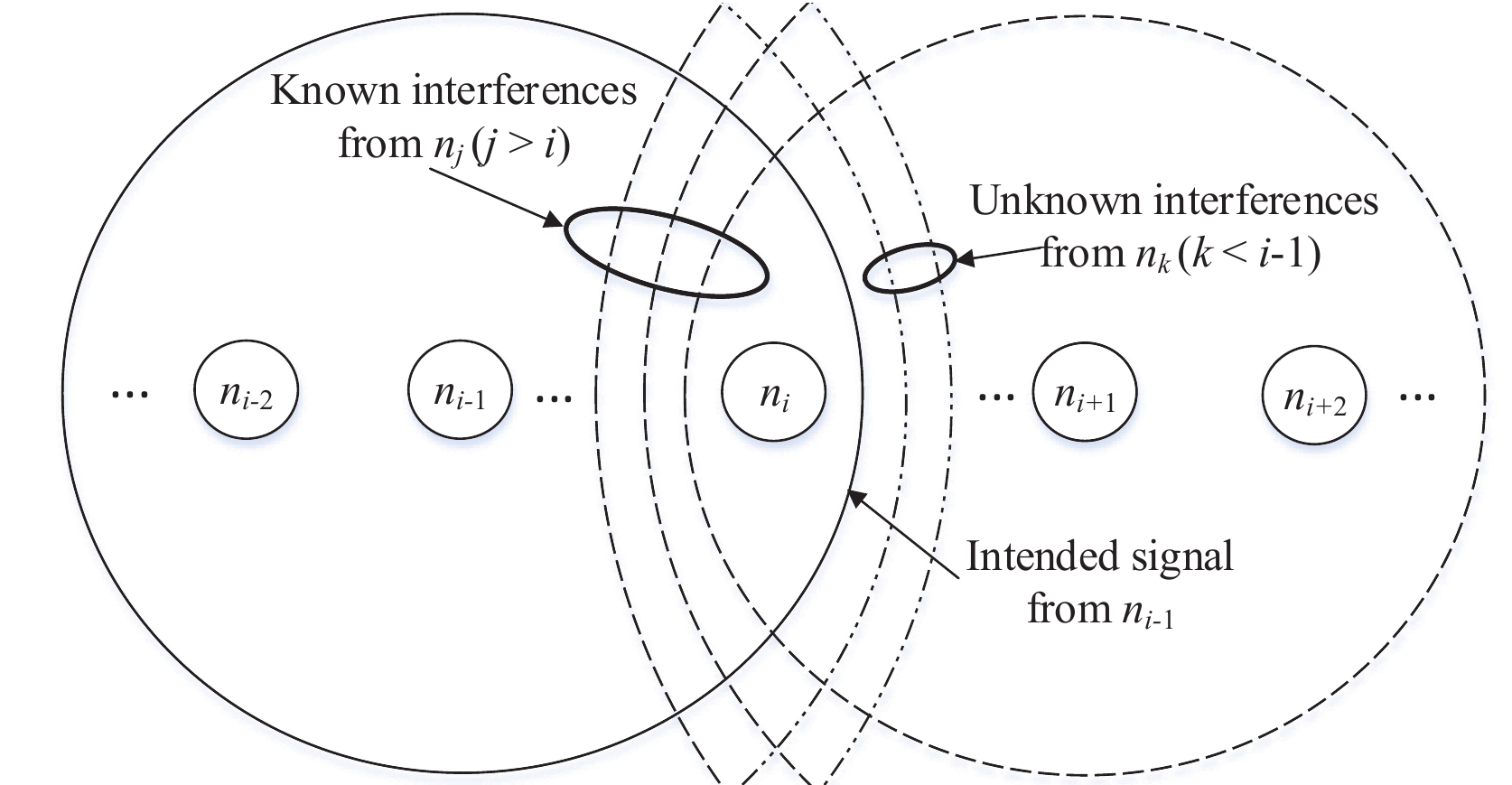}
\caption{Intended signal and interferences received by $n_{i}$ in a multi-hop flow with $N$ nodes, where E2E-KIC is used and nodes  simultaneously transmit their signals.}
\label{E2EKICInterference}
\vspace*{-1pt}
\end{figure}

As shown in Fig. \ref{E2EKICInterference}, when nodes are simultaneously transmitting their signals using E2E-KIC, node $n_{i}$ in this flow receives three different types of signals, including known interferences from $n_{j}~(j>i)$, unknown interferences from $n_{k}~(k<i-1)$, and its intended signal from $n_{i-1}$. According to \cite{fullDuplex}, the interference caused by self-transmission can be cancelled to the noise floor by using self-interference cancellation techniques. We define the noise at a node as $z_{\text{n}}$, which consists of residual self-interference and common noises. We define $\mathbf{S}^{'}_i=\{i-1,i,i+1\}$ and $\mathbf{S}_i=\{1,2,...,N\}$. Let $h_{j,i}~(j\in \mathbf{S}_i,~j\neq i)$ denote the channel gain from node $n_{j}$ to node $n_{i}$, and $x_{j}$ denote the signal sent by node $n_{j}$. Note that the average transmission power of node $x_{j}$ is equal to $E[|x_{j}|^2]$, where $E[|\cdot|]$ denotes the expected value. The signal received by $n_{i}$ can be written as
\begin{equation}\label{received_signal}
y_{n_{i}} = h_{i-1,i}x_{i-1}+h_{i+1,i}x_{i+1}+\sum_{j\in \mathbf{S}_i \backslash \mathbf{S}^{'}_i}h_{j,i}x_{j}+z_{\text{n}}\,,
\end{equation}
%where the transmission power is normalized to make the expression clearer.
where the last two components are the interferences from source nodes other than those in $\mathbf{S}^{'}_i$ and the noise at $n_{i}$, which are considered to be small enough to ensure the reliability of packet reception. We note that the known inferences in $\mathbf{S}_i$ can be eliminated using KIC, where knowledge of channel coefficients is not required when using the blind KIC technique \cite{BKIC}. The elimination of unknown interferences in $\mathbf{S}_i$ is left as future work, where we anticipate that a properly designed routing and power control method may be beneficial. We also note that when the path attenuation is large, these unknown interferences are unlikely to have a significant impact on the received signal quality. The first two components in Equation (\ref{received_signal}) represent a superposed signal. Because $n_{i}$ knows the content of $x_{i+1}$ which was forwarded by itself in the previous timeslot, it can successfully derive its wanted signal $x_{i-1}$ by cancelling the known interference from $n_{i+1}$. Therefore, E2E-KIC is feasible through our signal analysis.
\subsection{Achievable Throughput in Idealized Case}\label{TAnalysis}
Let $m_{j}$ denote the $j$th packet sent by $n_{1}$. For PR, PNC, FD, and E2E-KIC, they all need $N-1$ timeslots to finish the end-to-end transmission of the first packet $m_{1}$. For the subsequent packets, when PR is used, the packet $m_j~(j>1)$ arrives at the destination node in the $(N-1+3(j-1))$th timeslot. The reason is that, as shown in Fig. \ref{InvolvedRelayings}(a), the packet $m_j$ has to wait for at least $3$ timeslots to access the wireless channel in order not to interfere with the transmission of $m_{j-1}$. In other words, the shortest waiting time before sending a new packet with PR is $3$ timeslots. To forward $M$ packets to the destination node $n_{N}$, PR needs at least $(N-1+3(M-1))$ timeslots. Therefore, the achievable throughput of this flow with PR is $R_{\text{PR}} = \frac{M}{N-1+3(M-1)}$. Similarly, as shown in Fig. \ref{PNCandFD}(a) and Fig. \ref{InvolvedRelayings}(b), the shortest waiting times before sending a new packet when using PNC and E2E-KIC are $2$ and $1$ timeslot(s), respectively. Thus, the achievable throughputs of PNC and E2E-KIC are respectively $R_{\text{PNC}} = \frac{M}{N-1+2(M-1)}$ and $R_{\text{E2E-KIC}} = \frac{M}{N-1+(M-1)}$.

If using FD, the shortest waiting time before sending a new packet is different when the number of packets $M$ is odd or even. From \figurename~\ref{PNCandFD}(b), we can see that the shortest waiting time of $m_{j+1}$ ($j \in \{1,3,5,7,\cdots\}$) after the transmission of $m_{j}$ is $1$ timeslot, while the shortest waiting time of $m_{j+2}$ after the transmission of $m_{j+1}$ is $3$ timeslots. In other words, it needs $N-1+4((M+1)/2-1)=N+2M-3$ timeslots in total for the destination node to receive $M \in \{1,3,5,7,\cdots\}$ packets. When $M$ is an even number, it needs $N-1+2((M-1)-1)+1=N+2M-4$ timeslots. Therefore, the achievable throughput of FD is
\begin{IEEEeqnarray}{ll}
\label{E2ET-FD}
R_{\text{FD}}=&\left\{\begin{aligned}
&\frac{M}{N+2M-3},&M \in \{1,3,5,\cdots\} \\
&\frac{M}{N+2M-4},&M \in \{2,4,6,\cdots\}
\end{aligned}\right.
\end{IEEEeqnarray}

When $M\to \infty$, we have $R_{\text{E2E-KIC}}=1$, $R_{\text{FD}}=R_{\text{PNC}}=1/2$ and $R_{\text{PR}}=1/3$, i.e., $R_{\text{E2E-KIC}}:R_{\text{FD}}:R_{\text{PNC}}:R_{\text{PR}}=1:\frac{1}{2}:\frac{1}{2}:\frac{1}{3}$. It is important to note that when a flow only has single hop, the achievable throughput is also equal to $1$. This result indicates that with E2E-KIC, the unidirectional data flow over multiple hops has the potential to achieve the single-hop performance.
\section{Overview of E2E-KIC MAC}\label{secOverviewMAC}%
In this section, we give an overview of E2E-KIC MAC. We first discuss the basic principles in the design of E2E-KIC MAC. Then, we introduce how to satisfy these principles through scheduling, cross-layer cooperation, overhead reduction, and unnecessary contention avoidance.
\subsection{Basic Principles}
\label{sub:basicPrinciple}
We consider a general network scenario where nodes are connected in an arbitrary topology and there exist multiple data flows. The term ``data flow'' specifies the data that are transmitted from a specific source node to a specific destination node, through a specific path in the network. Each node is equipped with an omnidirectional antenna and all nodes share a single channel. The network stays in stationary state, i.e., the node locations and traffic flows do not change. Similarly to the IEEE 802.11 standard, all nodes contend for the wireless channel using the carrier sense multiple access with collision avoidance (CSMA/CA) mechanism. When obtaining the right to access the channel, the node (denoted by $n_{i}$) checks whether there is an opportunity to collaborate with other nodes and complete the transmission using E2E-KIC. If there is such an opportunity, it notifies its collaborating nodes to initiate an E2E-KIC transmission.
To successfully start and perform E2E-KIC in a distributed manner, the MAC protocol is designed according to the following principles:

\begin{enumerate}
\item To support multiple data flows, there should be a way to distinguish different data flows and determine which data flow should be served.

\item Before initiating an E2E-KIC process, a node should know in advance about which of its neighbors should be informed and how many nodes are involved.

\item To perform E2E-KIC, an initiating node $n_{i}$ should inform other involved nodes to participate in this packet exchange process. Nodes involving in the E2E-KIC process should also notify their neighbors (including those which are not involved in the E2E-KIC process) of the expected channel occupation time, to reserve the wireless channel.
%All this information should be obtained with a distributed method because E2E-KIC MAC is distributed.

%Nodes should be able to identify interferences with a distributed method.
\item Each node needs to be able to identify the origins and contents of interference signals with a distributed method, so that it can cancel the known interferences.
%An interference should be first identified by the node receiving it before performing KIC.
%Because multiple signals are simultaneously superposed at multiple nodes, we need a distributed method to ensure all nodes can identify their received interferences.

\item The number of control frames should be minimized. Different nodes should concurrently transmit control messages if possible.
%Control and ACK information is delivered among multiple nodes. To reduce the overhead of E2E-KIC MAC, it needs to minimize the number of control frames and increase the opportunities of the concurrent transmission of control and ACK frames.
\item A mechanism is needed to avoid unnecessary channel contention among nodes that transmit data belonging to the same data flow.
\end{enumerate}

\subsection{Flow Identification and Scheduling}
When multiple data flows exist in the network, each flow is assigned a unique identification (ID). These flow IDs can be directly assigned in small networks, in which each source-destination pair can be assigned a flow ID, or dynamically allocated by a control node, where the control node can be the macro-basestation in a wireless heterogeneous network \cite{lei2013challenges}. In this paper, we assume that the flow IDs are known to all the nodes.
Each node maintains a flow list, which records information on all the data flows passing though itself. Particularly, each entry in the flow list records the flow ID, the previous and next hops of the flow, and the total number of nodes involved in this flow.
Each node also maintains a sending buffer, which stores the packets that are waiting to be sent. Packet transmissions are scheduled according to the first-in first-out (FIFO) principle, subject to some additional constraints as will be discussed in Section \ref{sec:scheAlgo}. When node $n_i$ accesses the channel, it checks whether E2E-KIC can be performed for the data flow containing the packet it wants to send.
\subsection{Cross-Layer Cooperation}\label{NodeScheduling}
Principles 2, 3, and 4 in Section \ref{sub:basicPrinciple} can be satisfied through cooperation between the network, MAC, and physical layers.
\subsubsection{Cooperation between Network and MAC Layers}
Principle 2 can be satisfied by allowing E2E-KIC MAC to obtain information from the network layer. When node $n_1$ wants to send packets to node $n_N$, the intermediate nodes (relays) $n_2,n_3,...,n_{N-1}$ can be found from the routing information in the network layer.
We then know the next and previous hops of each involved node, and also the total number of involved nodes ($N$). With this information, E2E-KIC MAC can operate in a way to satisfy Principle 3, as will be discussed in details in Section \ref{sectMAC}. Note that each node does \emph{not} need to have knowledge about the complete routing table. The appendix of our technical report \cite{OurTechRep} shows an example how the necessary information can be obtained in the ad hoc on-demand distance vector (AODV) protocol.

In the remaining discussion, we define the node set $\mathbf{F} \triangleq \{n_{1},n_{2},\cdots,n_{N}\}$ as the path of a specific data flow, and the nodes in $\mathbf{F}$ form a chain topology with $N$ nodes, which is a sub-network of the original network. We will use $\mathbf{F}$ to denote both the flow itself and the nodes contained in the flow path.
%To satisfy Principle 1, nodes in $\mathbf{F}$ are scheduled in an E2E-KIC process. According to this scheduling strategy, the neighbors that should be informed by node $n_{i}$ are its next \emph{and} previous hops, while the neighbor of $n_{j}$ $(j\neq i, 1<j<N)$ in $\mathbf{F}$ that should informed is its next \emph{or} previous hop.
%Therefore, through the cooperation between MAC and network layers, nodes can know in advance which neighbor(s) should be informed and the number of involved nodes with a distributed method.

\subsubsection{Cooperation between MAC and Physical Layers}
To satisfy Principle 4, we utilize the MAC header to identify different packets. E2E-KIC MAC does not superpose the header parts of data signals, so that the headers of each independent packet can be decoded. In order to ensure that the headers are not superposed, the bit sequence of transmitted packets are reversed when necessary. After identifying the packets, the KIC operation is performed on physical-layer signals to extract the wanted packet from the superposed signal.
%This purpose is achieved through selecting the suitable bit sequence and transmitted time for the data frame in the MAC layer. Readers can refer to Section \ref{E2ETiming} for details about the selection method. After being identified in the physical layer, the interference in a superposed signal can be cancelled with KIC if its content is known through searching the cached packets that has been successfully transmitted.

\subsection{Overhead Reduction}\label{ControlFrameExchange}
%Overhead reduction is necessary to maintain the performance of a MAC protocol. In this subsection, we discuss how to reduce the overhead of E2E-KIC MAC through decreasing the exchange time of the control and ACK frames. The main methods can be summarized as minimizing the number of frames and concurrent transmission through space reuse and KIC.
The control overhead is maintained to be low by limiting the number of control frames to be transmitted and allowing concurrent transmissions through spatial reuse and KIC.

\subsubsection{Exchange Strategy of RTS/CTS Frames}
Because the MAC is distributed, any node $n_i\in\mathbf{F}$ may get the right to access the channel and initiate an E2E-KIC transmission from $n_1$ to $n_N$.
Node $n_i$ sends an RTS packet to notify its neighbors of the E2E-KIC opportunity and to reserve the channel. Its neighbors then propagate this information with CTS packets. Each node in $\mathbf{F}$ only sends one control packet before the E2E-KIC process starts.

%In the considered scenario, node $n_{i}$ is the source of the scheduling information (such as, which a flow is scheduled, the number of involved nodes, etc.). This information should be spread from $n_{i}$ to end nodes (i.e., $n_{1}$ and $n_{N}$) through the control frame. Because the wireless channel has the nature of broadcasting, through once transmission of the control frame, a node can achieve at most three purposes: 1) responding the scheduling notification from its neighbor; 2) informing its corresponding neighbor in $\mathbf{F}$ to participate in this scheduling; 3) reserving the channel to avoid unnecessary contentions. Therefore, it only needs once control-frame transmission for a node to spread the scheduling information and reserve the channel in E2E-KIC MAC. Based on this observation, we propose a exchange strategy for the control frame, which has a '$<$'-shaped timing diagram as shown in Fig. \ref{timingdiagram}. With this strategy, one node transmits one control frame in an E2E-KIC process, and at the same time, two nodes simultaneously transmit their control frames as pair if the space resource can reused.
\subsubsection{Exchange Strategy of ACK Frames}
A node should acknowledge the successful reception of the data frame sent by its previous hop. Two adjacent nodes can transmit ACK frames simultaneously because nodes are in the FD mode.

\subsection{Unnecessary Contention Avoidance}\label{sec:scheAlgo}
When node $n_i$ and node $n_j$ both plan to send packets that belong to the same flow $\mathbf{F}$ in their next channel access, only one node needs to contend for the channel and initiate an E2E-KIC transmission, so that both nodes can transmit one packet after the contending node obtains the channel access. Such an operation is more efficient than having multiple nodes belonging to $\mathbf{F}$ contend for the channel, because the latter option would cause unnecessary channel contentions which may result in low channel utilization. To make the channel contention procedure adaptable to network conditions, we design a distributed access-control method, which consists of two components:
\begin{itemize}
\item Contention window size adaptation, which ensures that nodes with heavy-loaded buffer have higher access probabilities;
\item Contention reduction, which restricts the transmissions of nodes under some conditions.
\end{itemize}
In the following, we introduce the access-control method and its impact to packet queuing/scheduling.

\subsubsection{Contention Window Size Adaptation}
The contention window size indicates the access probability of nodes. The main idea of contention window size adaptation is tuning the contention window size (thus the access probability) of nodes as a function of their historical channel access frequencies. Readers can refer to \cite{EFFD} for more details.

\subsubsection{Contention Reduction}
The basic idea of contention reduction is that for each flow $\mathbf{F}$, half of the nodes in $\mathbf{F}$ should wait for an additional time $T_{\text{wait}}$ before contending for the channel.
%Let $\alpha$ denote the index (starting from $1$) of a node in $\mathbf{F}$.
When node $n_i$ receives a new packet belonging to flow $\mathbf{F}$, node $n_i$ sets a timer to wait for $T_{\text{wait}}$ when one of the following conditions is satisfied: 1) $\mathbf{F}$ contains an odd number of nodes and $i$ is an even number; 2) $\mathbf{F}$ contains an even number of nodes and $i$ is an odd number. The reason for the above conditions is mainly considering that the destination node does not need to send packets. Each node $n_i$ maintains a timer for each flow $\mathbf{F}$ where $n_i \in \mathbf{F}$, and it only contends for the channel to send packets belonging to flow $\mathbf{F}$ when the timer is cleared. With this method, excessive contention among nodes that intend to transmit packets belonging to the same flow is reduced.

The introduction of $T_{\text{wait}}$ also impacts the packet transmission sequence at each node. Node $n_{i}$ sends the first packet in its sending buffer which is not under the waiting status, and the node attempts to initiate E2E-KIC for the flow to which this packet belongs.
\section{E2E-KIC MAC Protocol}\label{sectMAC}
\begin{figure*}[!t]
\centering \includegraphics[width=0.99\textwidth]{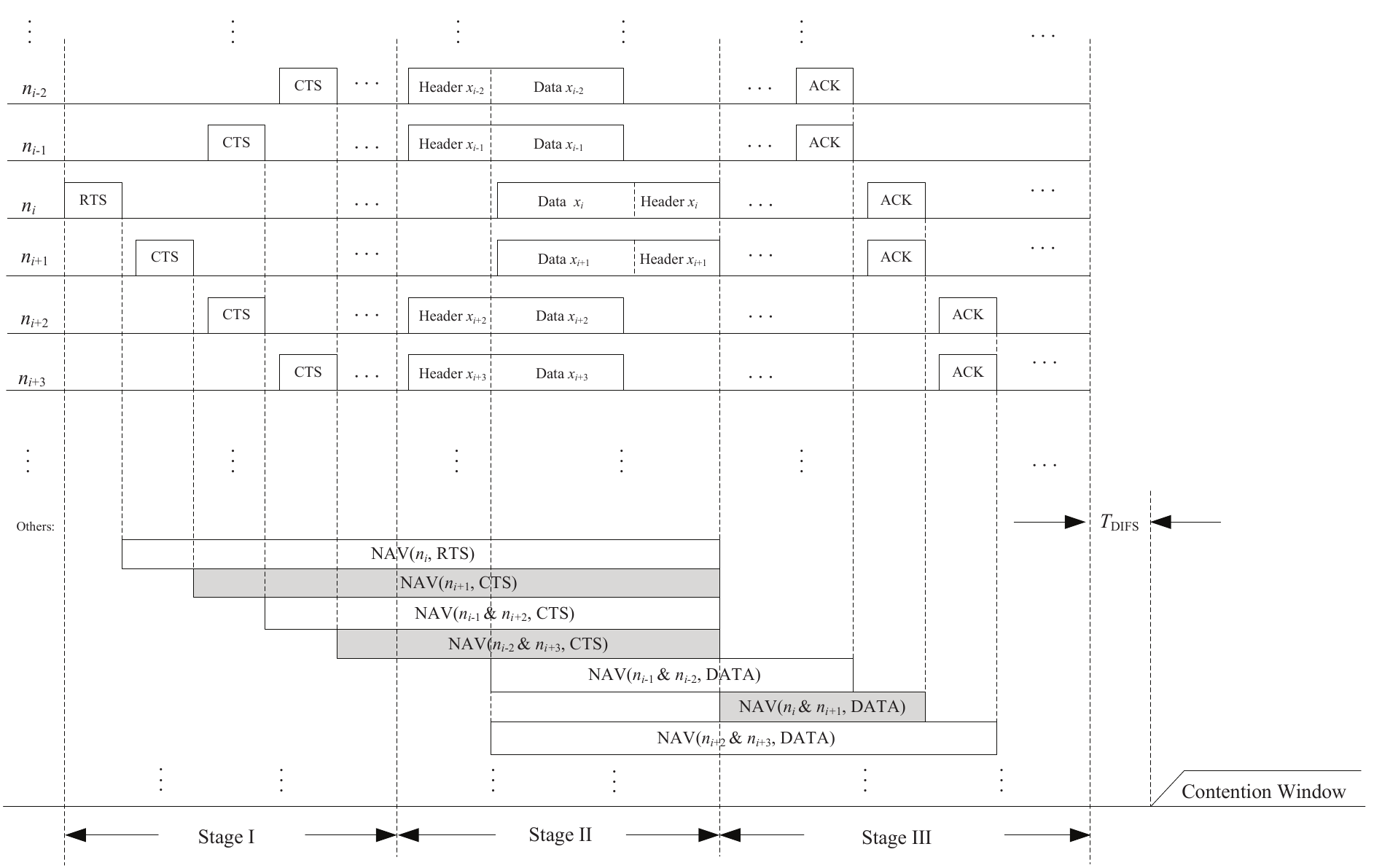}
\caption{Standard timing diagram of packet exchange in E2E-KIC MAC, when $n_{i}$ initiates an E2E-KIC process.}
\label{timingdiagram}
\vspace*{-2pt}
\end{figure*}

In this section, we present the details of E2E-KIC MAC, in particular its timing procedure. The proposed E2E-KIC MAC is an extension of the RTS/CTS-based IEEE 802.11 MAC protocol. We consider a data flow $\mathbf{F}$ with $N$ nodes, and these $N$ nodes (which are connected as a chain) are part of a bigger multi-hop network.
%The data flow is from $n_{1}$ to $n_{N}$.
Unless specifically stated, we restrict our discussion to the nodes in flow $\mathbf{F}$ in this section.
We assume that node $n_{i}\in\mathbf{F}$ has obtained the right to access the wireless channel, as shown in \figurename~\ref{timingdiagram}.

We define $n_{j}$ $(j<i)$ as an anterior node of $n_{i}$, and $n_{k}$ $(k> i)$ as a posterior node of $n_{i}$. The number of the involved anterior (or posterior) nodes is defined as the anterior-hop (or posterior-hop) limit. Let $T_{\text{SIFS}}$ and $T_{\text{DIFS}}$ denote the short inter-frame space (SIFS) and the distributed inter-frame space (DIFS) as in the IEEE 802.11 standard \cite{M80211}. Let $T_{\text{PHY-Hd}}$, $T_{\text{MAC-Hd}}$, $T_{\text{CTS}}$, $T_{\text{DATA}}$, and $T_{\text{ACK}}$ respectively denote the time length of physical-layer header, MAC-layer header, CTS, data part of data frame, and ACK. We assume that data packets have the same lengths to simplify our discussion.
\subsection{Timing of E2E-KIC MAC}\label{E2ETiming}
As shown in Fig. \ref{timingdiagram}, there are three stages in the packet exchange procedure of E2E-KIC MAC. In Stage \uppercase\expandafter{\romannumeral1}, the RTS frame\footnote{Frame formats are introduced in our technical report \cite{OurTechRep}.} initiates the packet exchange process of E2E-KIC MAC, and the CTS frames inform other nodes to get ready for packet exchange. In Stages \uppercase\expandafter{\romannumeral2} and \uppercase\expandafter{\romannumeral3}, data packets are exchanged and receptions are acknowledged by receiving nodes. The details of the three stages are described in the following.

\subsubsection{RTS/CTS Stage} When $n_{i}$ has the right to access the channel, an RTS frame is broadcasted to all the neighbors of $n_{i}$ to reserve the wireless channel. The RTS frame contains the anterior and posterior hop limits. These fields can be utilized by hop-limit decision algorithms with the purpose of achieving the trade-off of transmission efficiency and collision probability. The design of specific hop-limit decision algorithms is left for future work. In the proposed E2E-KIC MAC, the anterior and posterior hop limits are respectively set to the number of all the anterior and posterior nodes (seen from node $n_i$) of the scheduled flow. When the anterior-hop limit is greater than $1$, the address of $n_{i-1}$ is also included in the RTS frame in order to inform $n_{i-1}$. Besides, the RTS frame also contains the ID of the flow that will be scheduled. With flow ID and hop limits, the nodes receiving the RTS frame can know which neighbor(s) should be informed.
%There is also a hop-count field in the RTS frame (denoted by $H_{\text{RTS}}$), which is set to zero \textcolor{red}{(Why do we need hop count in RTS if it is always set to zero?)}. The hop count denotes how many hops the information of the scheduled data flow (i.e., the flow ID and the hop limits) has been forwarded. With flow ID, hop count, and hop limits, the nodes receiving the RTS frame can know which neighbor(s) should be informed.

After receiving the RTS frame, node $n_{i+1}$ responds with a CTS frame to $n_{i}$ after time $T_{\text{SIFS}}$, while node $n_{i-1}$ responds with a CTS frame after time $2T_{\text{SIFS}}+T_{\text{CTS}}$. This time difference ensures that the two CTS frames are not superposed at $n_{i}$, because otherwise $n_i$ cannot decode either of them and does not know whether the RTS frame has been successfully received by both of its neighbors. There is a hop-count field in the CTS frame (denoted by $H_{\text{CTS}}$), which records how many hops the scheduling information has been forwarded. The hop counts in the two CTS frames sent by nodes $n_{i-1}$ and $n_{i+1}$ are both set to $H_{\text{CTS}}=1$. The anterior-hop limit, posterior-hop limit, and the flow ID fields in the CTS frame are set to the same values as those in the RTS frame. A CTS frame also has at most two receiver addresses, which can be derived from information in the flow list according to the flow ID. The first receiver address $\text{RA}_{1}$ of the CTS frame sent by $n_{i-1}$ (or $n_{i+1}$) is set to the address of $n_{i}$. The second receiver address $\text{RA}_{2}$ of the CTS frame sent by $n_{i-1}$ (or $n_{i+1}$) is set to the address of $n_{i-2}$ (or $n_{i+2}$), if the anterior-hop (or posterior-hop) limit is not smaller than $H_{\text{CTS}}+1$ (i.e., the hop count in the CTS frame sent by $n_{i-1}$ (or $n_{i+1}$) plus one).

In general, when node $n_{j}$~$(j<i)$ (or $n_{k}$~$(k>i)$) receives a CTS frame from node $n_{j+1}$ (or $n_{k-1}$), it responds with a CTS frame after time $T_{\text{SIFS}}$ if the hop count is within the limit as discussed earlier. The hop count in the new CTS frame increments by $1$ compared with the hop count in the CTS frame that the node has received. When node $n_{j}$~$(j<i)$ (or $n_{k}$~$(k>i)$) receives a CTS frame from any node \emph{other than} $n_{j+1}$ (or $n_{k-1}$), it keeps silent (i.e., does not send any packet) until Stage \uppercase\expandafter{\romannumeral2}. As shown in \figurename~\ref{timingdiagram}, with this sending sequence, two nodes  $n_{i-j}$ and $n_{i+j+1}$ ($0<j<\min(i,N-i)$, where $\min(x,y)$ denotes the minimum between $x$ and $y$) simultaneously transmit their CTS frames as a pair because they do not interfere with each other, which limits the control overhead.

\subsubsection{Data Exchange Stage} The basic principle of the design of Stage \uppercase\expandafter{\romannumeral2} is to make sure that the superposed data frames can be successfully decoded. As shown in Fig. \ref{timingdiagram}, the involved nodes simultaneously enter Stage \uppercase\expandafter{\romannumeral2} after exchanging control frames. Any two nodes with two-hop distance transmit data frames alternately with normal bit sequence and adverse bit sequence (i.e., the tail of the data frame is sent first).

The time that a nodes enters Stage \uppercase\expandafter{\romannumeral2} is calculated in a distributed manner according to the time of receiving a particular CTS frame and the hop count recorded in that CTS frame. The formulas for calculating the time differ slightly among different sets of nodes, and they are summarized as follows. Let $A$ and $P$ respectively denote the anterior-hop limit and the posterior-hop limit.
\begin{itemize}
\item For node $n_l$ with $i+2\leq l \leq N$ (recall that $n_i$ is the node sending the RTS frame), after receiving the CTS frame from node $n_{l-1}$, node $n_{l}$ enters Stage \uppercase\expandafter{\romannumeral2} after time
\begin{equation}
\label{DataSendTime3}
T_\textrm{before-State-II}=(\max(A+1,P)-H_{\text{CTS}})(T_{\text{CTS}}+T_{\text{SIFS}})\,,
\end{equation}
where $\max(x,y)$ denotes the maximum between $x$ and $y$. The expression $\max(A+1,P)$ represents the total number of CTS-exchange timeslots (i.e., $T_{\text{CTS}}+T_{\text{SIFS}}$). The anterior-hop limit $A$ is added by one because node $n_{i-1}$ has to wait for an additional CTS-exchange timeslot before sending its CTS frame, as shown in Fig. \ref{timingdiagram}. The expression $(\max(A+1,P)-H_{\text{CTS}})$ denotes the number of remaining CTS-exchange timeslots.

\item For nodes $n_{i-1}$, $n_{i}$, and $n_{i+1}$, after sending (for $n_{i}$) or receiving (for $n_{i-1}$ and $n_{i+1}$) the RTS frame, they enter Stage \uppercase\expandafter{\romannumeral2} after time $T_\textrm{before-State-II}=\max(A+1,P)(T_{\text{CTS}}+T_{\text{SIFS}})$.

\item For node $n_l$ with $1\leq l \leq i-2$, after receiving the CTS frame from node $n_{l+1}$, node $n_{l}$ enters Stage \uppercase\expandafter{\romannumeral2} after time $T_\textrm{before-State-II}=(\max(A+1,P)-H_{\text{CTS}}-1)(T_{\text{CTS}}+T_{\text{SIFS}})$ for the same reason that an additional CTS-exchange timeslot is needed for node $n_{l-1}$.
\end{itemize}

In the following, we discuss the bit sequence. We first define an indexing variable $\alpha_l = l-(i-A)+1$ to simplify our discussion.
%Node $n_{N}$ does not send packets at Stage \uppercase\expandafter{\romannumeral2}, because it is the destination node of the scheduled flow.
Let $\beta$ denote the bit sequence of a packet sent by node $n_l$, the bit-reversing specification of E2E-KIC MAC can be formally expressed as
\begin{equation}
\beta=\mathrm{mod_2}(((2\alpha_l+1-\mathrm{sign}(\mathrm{mod_2}(\alpha_l+1)-0.5))/4))\,,
\end{equation}
where $\mathrm{mod_2}(\cdot)$ is the modulo 2 function and $\mathrm{sign}(\cdot)$ is the sign function. The value $\beta = 1$ means that the bit sequence of a packet is normal, and the packet is sent after time $T_{\text{SIFS}}$ when entering Stage \uppercase\expandafter{\romannumeral2}. The value $\beta = 0$ indicates that a packet is sent with the adverse bit sequence (i.e., the tail of the data frame is sent first), after a time of $T_{\text{frame-diff}}+T_{\text{SIFS}}$ after entering Stage \uppercase\expandafter{\romannumeral2}, where we define $T_{\text{frame-diff}}=T_{\text{SIFS}}+T_{\text{PHY-Header}}+T_{\text{MAC-Header}}$. This ensures that only the data parts of two frames from $n_{l-1}$ and $n_{l+1}$ are superposed at $n_{l}$ $(2\leq l \leq N-1)$. Because the interference caused by self-transmission of $n_{l}$ can be effectively cancelled, the headers of both frames from $n_{l-1}$ and $n_{l+1}$ can be decoded, which can be used to identify both superposed frames.
\subsubsection{ACK Exchange Stage} A node sends an ACK frame to its previous hop if successfully receiving a packet. Because nodes are in full-duplex mode, two adjacent nodes can simultaneously send their ACK frames, which reduces the time for ACK exchange. After the end of the data exchange process, node $n_l$ responds with an ACK frame after time
\begin{IEEEeqnarray}{lll}
T_\textrm{before-ACK}&{}={}&(((2\alpha_l-1-\mathrm{sign}(\mathrm{mod_2}(\alpha_l)-0.5))/4) \nonumber \\
{}&{}&-1)T_{\text{ACK}}+(((2\alpha_l-1-\mathrm{sign}(\mathrm{mod_2}(\alpha_l) \nonumber \\
{}&{}&-0.5))/4))T_{\text{SIFS}},
\end{IEEEeqnarray}
where $\alpha_l\in [2,A+P+1]$ because the first node in the E2E-KIC process $n_{i-A}$ does not need to send ACK, and there are $A+P+1$ nodes involved in the E2E-KIC process in total.
\subsection{Exception Handling}\label{Exce_handling}
Section \ref{E2ETiming} has presented the basic timing procedure of E2E-KIC MAC. However, there are cases where we cannot follow the basic timing, for instance if there is a frame loss, if a node has no data to send, etc. In the following, we discuss how E2E-KIC MAC handles these cases.

\subsubsection{CTS Loss} A node $n_l$ does not send data packets if it has not received the CTS frame from its next hop $n_{l+1}$. If node $n_{l}$ does not receive the CTS frame from its previous hop $n_{l-1}$ but successfully receives its wanted data packet, it still sends the ACK frame.

\subsubsection{No Data to Send} When node is informed to participate in the E2E-KIC process but has no packet to send, it still sends a CTS frame to  inform its neighboring nodes (subject to the hop limit constraint). This ensures that other nodes belonging to the same flow path can be informed.

\subsubsection{Data Packet Error or Cannot Be Decoded} If a superposed data packet is erroneous or cannot be decoded because information that is necessary for performing KIC is missing, the node does not send ACK and waits for the end of the packet exchange process. The waiting is necessary because, otherwise, other transmissions in this E2E-KIC process may be interfered.

\subsubsection{ACK Loss} When a node does not receive the corresponding ACK frame after sending a data packet, it will retransmit the data packet in the next channel access unless the number of the retransmissions exceeds its limit.

\subsection{Network Allocation Vector (NAV) Setting}
\label{NAV_setting}
The NAV is used for channel reservation. Every time when a frame arrives at a node that does not participate in the E2E-KIC process, the node updates its NAV and remains silent until the channel reservation time specified by the NAV expires.

The RTS frame requests channel occupation until the end of Stage \uppercase\expandafter{\romannumeral2}. Formally, the NAV length of the RTS frame is $T_{\text{NAV}}(\text{RTS}) = (\max(A+1,P))(T_{\text{SIFS}}+T_{\text{CTS}})+T_{\text{DATA}}+2T_{\text{frame-diff}}$.

%\begin{IEEEeqnarray}{ll}\label{NAV_RTS}
%T_{\text{NAV}}(\text{RTS}) =& (\max(A+1,P))(T_{\text{SIFS}}+T_{\text{CTS}})
%+T_{\text{DATA}}+2T_{\text{frame-diff}}\,.
%\end{IEEEeqnarray}

The NAV length of CTS is set to ensure that the neighbors of the involved nodes keep silent before the end of Stage \uppercase\expandafter{\romannumeral2}. The NAV length of CTS sent by node $n_{j}$~$(j<i)$ (or $n_{k}$~$(k>i)$) is calculated according to the NAV length of the RTS/CTS frame received from $n_{j+1}$ (or $n_{k-1}$). Specifically, the NAV length of node $n_{l}$ is calculated by
\begin{equation}
T_{\text{NAV}}(\text{CTS},n_{l}) =\begin{cases}
T_{\text{NAV}}(\text{CTS},n_{l+1})-\theta & \textrm{if }l<i-1\\
T_{\text{NAV}}(\text{RTS})-2\theta & \textrm{if }l=i-1\\
T_{\text{NAV}}(\text{RTS})-\theta & \textrm{if }l=i+1\\
T_{\text{NAV}}(\text{CTS},n_{l-1})-\theta & \textrm{if }l>i+1\\
\end{cases}
\label{CTSNAV}
\end{equation}
where $\theta=T_{\text{CTS}}+T_{\text{SIFS}}$.
%\begin{equation}\label{NAV_CTS_1}
%T_{\text{NAV}}(\text{CTS},n_{i-1}) = T_{\text{NAV}}(\text{RTS})-2(T_{\text{CTS}}+T_{\text{SIFS}})\,,
%\end{equation}
%\begin{equation}\label{NAV_CTS_2}
%T_{\text{NAV}}(\text{CTS},n_{i+1}) = T_{\text{NAV}}(\text{RTS})-(T_{\text{CTS}}+T_{\text{SIFS}})\,.
%\end{equation}
%In other cases, the NAV length of $n_{j}$ $(j<i-1)$ is calculated by
%\begin{equation}
%T_{\text{NAV}}(\text{CTS},n_{j})= T_{\text{NAV}}(\text{CTS},n_{j+1})-(T_{\text{CTS}}+T_{\text{SIFS}})\,.
%\end{equation}
%The NAV length of $n_{k}$ $(k>i+1)$ is
%\begin{equation}
%T_{\text{NAV}}(\text{CTS},n_{k})=T_{\text{NAV}}(\text{CTS},n_{k-1})-(T_{\text{CTS}}+T_{\text{SIFS}})\,.
%\end{equation}

The NAV length of a data frame is set to cover the remaining time from decoding its header until receiving its intended ACK frame. This procedure is similar as in \cite{MAC}. Specifically, the NAV length of the data frame sent by node $n_l$ is set to
\begin{IEEEeqnarray}{ll}
T_{\text{NAV}}(\text{DATA})&{}= ((2\alpha_l+1-\text{sign}(\mathrm{mod_2}(\alpha_l+1)-0.5))/4) \nonumber \\
&(T_{\text{SIFS}}+T_{\text{ACK}})+\beta T_{\text{frame-diff}}\,.
\end{IEEEeqnarray}
%\subsection{Frame Formats}
%This section introduces the frame formats in the proposed E2E-KIC MAC protocol. We modify the frames defined in the IEEE 802.11 standard to enable additional functionalities, as shown in \figurename~\ref{frameFormats}.
%\subsubsection{RTS}
%Different from IEEE 802.11, the RTS frame in E2E-KIC MAC contains two receiver addresses and a three-byte control information. The three-byte control information includes the ID of the flow scheduled in this round, and the anterior and posterior hop limits to control the number of involved nodes.
%\subsubsection{CTS}
%The CTS frame has the similar format as the RTS frame except for the hop-count field (which records the number of hops that the scheduling information has passed through). The reason is that similarly to the RTS frame, the CTS frame is used to spread the scheduling notification, and at the same time, respond to this notification.
%\subsubsection{Data}
%The data frame in E2E-KIC MAC is similar to that in IEEE 802.11, except that it contains a flow-ID field to identify the flow that the packet belongs to.
%
%\begin{figure}[!t]
%\centering \includegraphics[width=0.8\textwidth]{picture/frame_formats.pdf}
%\caption{Frame formats of the E2E-KIC MAC protocol.}
%\label{frameFormats}
%%\vspace*{-3pt}
%\end{figure}
\section{Throughput Analysis} \label{secThroughputAna}
This section focuses on analyzing the throughput of the E2E-KIC MAC protocol. It is a challenging job to analyze the throughput of wireless multi-hop networks \cite{Bianchi1,ANCMAC,ng2007throughput,jang2012ieee,tsertou2005new,ThroughputTingChao}. Compared with existing work \cite{Bianchi1,ANCMAC,ng2007throughput,jang2012ieee,tsertou2005new,ThroughputTingChao}, our analysis is more challenging because: 1) we consider multi-hop flows and hidden terminals, 2) and multiple concurrent transmissions in E2E-KIC lead to stronger interdependence among links.

We aim to analyze the throughput in the saturation condition, i.e., a node always has packets to send when it has right to access the channel or is cooperated to participate in an E2E-KIC process. The analysis approach is briefly summarized as follows. As in \cite{Bianchi1}, we assume that each node has a state, which is represented by a variable equal to the number of remaining backoff timeslots. The state transition of a node is modeled as a Markov chain, where a node at a considered timeslot can either initiate an E2E-KIC process with probability $P_{\text{t}}$, or be cooperated by other nodes to participate in an E2E-KIC process with probability $P_{\text{c}}$. For simplicity, we assume in our analysis that the initiation events are independent and identically distributed (i.i.d.) among all nodes, and we also assume that the cooperation events are i.i.d. among all nodes. With this assumption, we first find a relationship between $P_{\text{t}}$ and $P_{\text{c}}$ by analyzing the stationary distribution of the Markov chain. Then, we find another relationship between $P_{\text{t}}$ and $P_{\text{c}}$ by analyzing the channel-sharing principles of E2E-KIC MAC. We find the saturation throughput expression in the end.
\subsection{Initiation Probability}
For a given node, let $s(t)$ and $b(t)$ respectively denote the two stochastic processes representing the sizes of the contention window and the back-off counter at time $t$. According to \cite{Bianchi1}, the stochastic process $\{s(t), b(t)\}$ is a two-dimensional Markov chain.
However, it is very difficult to analyze the dynamics of the contention window size in E2E-KIC MAC, because of the complicated backoff algorithm in E2E-KIC MAC (as presented in Section \ref{sec:scheAlgo}). For simplicity, we assume that the size of the contention window ($W$) is constant in our analysis. We can see from the simulations in Section \ref{sectSim:gridTopo} that the impact of contention window size is insignificant in our considered topology (presented below), and we leave more comprehensive analysis as future work. Based on the stationary distribution of the Markov chain $\{b(t)\}$, we can express the initiation probability as
\begin{equation}
P_{\text{t}}=\frac{Z P_{\text{c}}}{ZWP_{\text{c}} + \left(1-P_{\text{c}}\right)\left(W-Z\right)} \,,
\label{TimeDomiEqn}
\end{equation}
where $Z=\left(1-P_{\text{c}} \right)^{-W} / ({1-\left(1-P_{\text{c}}\right)^{-1}})$. More details about the derivation of  (\ref{TimeDomiEqn}) can be found in our technical report \cite[Section VI-A]{OurTechRep}. Equation (\ref{TimeDomiEqn}) expresses a relationship between $P_{\text{t}}$ and $P_{\text{c}}$. To calculate these probabilities, we still need another relationship between them.
\subsection{Cooperation Probability}\label{sec:coopProba}
We derive the expression of $P_{\text{c}}$ according to the channel-sharing principles of E2E-KIC MAC as follows. Some assumptions are made to simplify our analysis. Based on these assumptions, $P_{\text{c}}$ is expressed as a function of $P_{\text{t}}$.
\subsubsection{Assumptions}
\begin{figure*}[!t]
	\centerline{\subfigure[]{\includegraphics[width=0.365\textwidth]{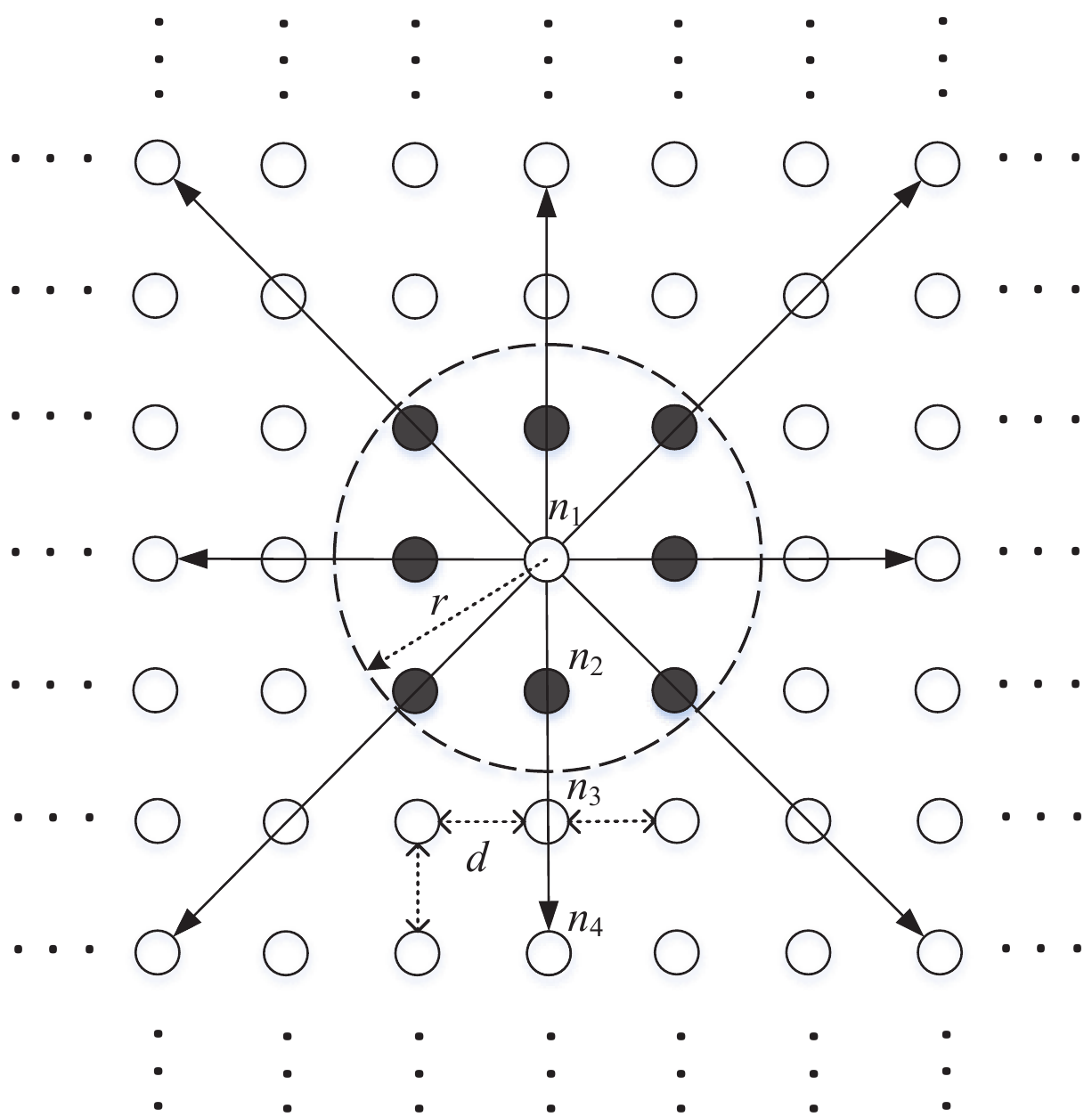}
			\label{figFlowConfiguration}}
		\hfil
		\subfigure[]{\includegraphics[width=0.4\textwidth]{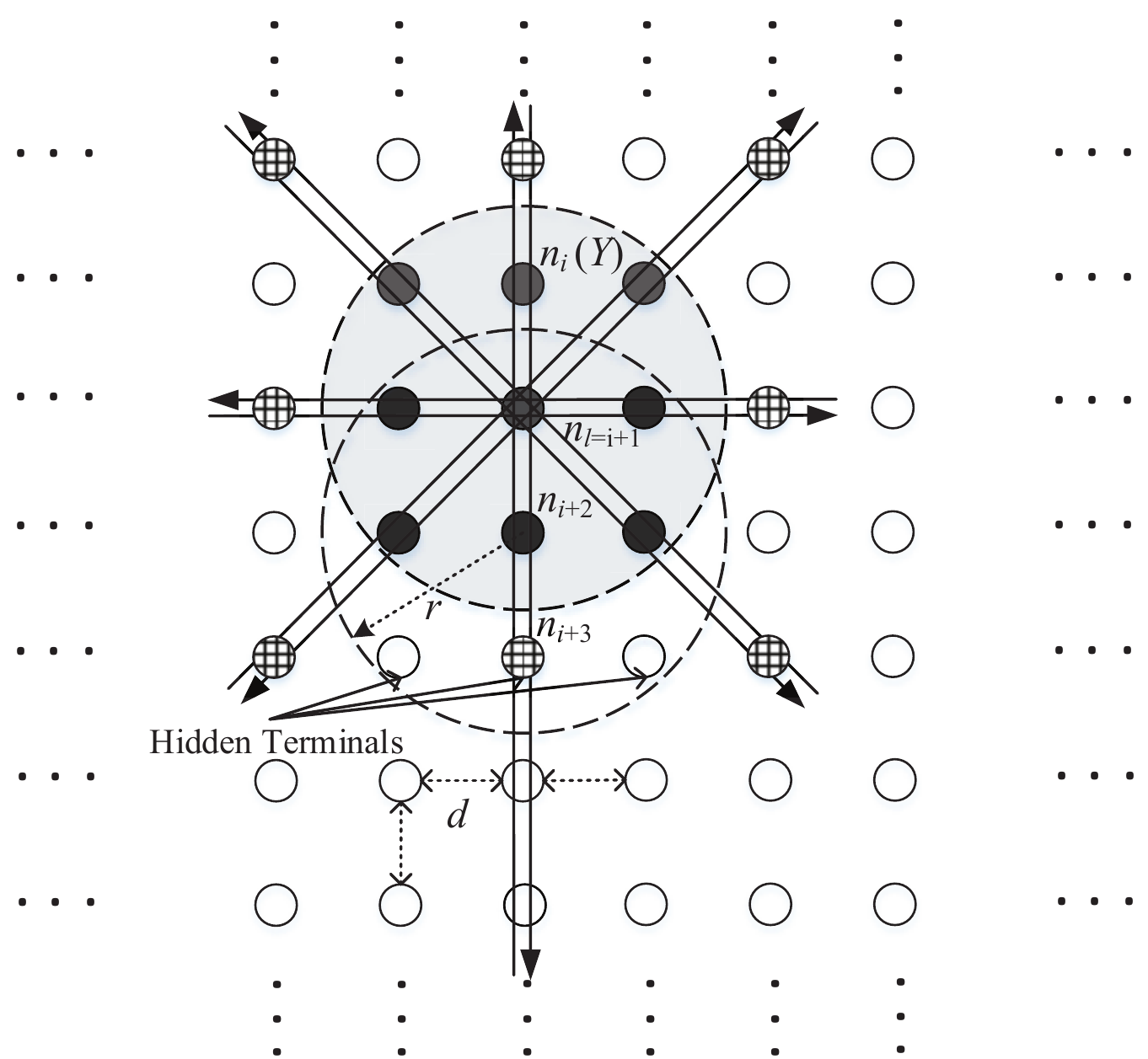}
			\label{figAnE2EKICProcess}}}
	\caption{An example of flow configurations at a two-dimensional grid network: (a) flows with $n_{1}$ as the source node, where the path size $N_{\text{path}}$ is equal to four; (b) flows with $n_{i+1}$ as their $(l=i+1)$th cooperator. Variable $d$ denotes the distance between two nodes in the same row or column. Variable $r$ denotes the radius of the CS or transmission range. The big cycles with dashed line show the CS or transmission ranges of nodes $n_{i+1}$ and $n_{i+2}$. An example node $n_{i+1}$ can be $(i+1)$th cooperator of eight flows (which originate at nodes marked with grid background in the figure), because it has eight neighbors.}
	\label{figRegurlarMeshGrid}
	\vspace*{-2pt}
\end{figure*}
Due to the complexity of our problem, similarly to existing work in \cite{Bianchi1,ANCMAC,jang2012ieee,ng2007throughput,tsertou2005new,ThroughputTingChao}, we make a few assumptions in our analysis to make the problem mathematically tractable. As shown in Fig. \ref{figRegurlarMeshGrid}, we consider a two-dimensional grid network with an infinite number of nodes, where the distance $d$ between any two adjacent nodes in the same row or column is constant. To simplify our analysis, we assume that the carrier sense (CS) range of a node is the same as its transmission range. The analysis of a more realistic scenario, for an arbitrary multi-hop topology and CS range, can also be conducted with a similar approach as what we use in our paper, but with a larger number of events that can lead to state transition. Due to space limitation, we leave this analysis to future work. We also note that the insights derived from the grid network should also apply in arbitrary networks, because when node locations in arbitrary networks obey uniform distribution, the node distribution is similar to that of a grid network\footnote{This means when considering a specific area of the network, the number of nodes under arbitrary networks with uniform node distribution is similar to that under the grid topology.}. It follows that we expect similar performance trends in these two types of networks. The performance in arbitrary networks is demonstrated in the simulations in Section \ref{RandomTopologySim}.

Let $r$ denote the radius of the CS range. We set $d \leq r < 2d$ to ensure that a node can only interfere or communicate with its neighbors. Based on the above assumptions, the CS range of a node only covers the node and its neighbors, where the number of neighbors can be either four or eight (see Fig. \ref{figRegurlarMeshGrid}). To ignore the physical-layer issues, the protocol interference model is assumed, i.e., if two nodes are within the transmission ranges of each other, the path gain between them is equal to $1$; otherwise, the path gain is equal to $0$ \cite{CapacityGupta}. Let $N_{\text{CS}}$ denote the number of nodes within the CS range of a node (including the node itself). Based on this interference model, the considered topology only has two different cases: 1) $N_{\text{CS}}=9$, if $\frac{r}{2} < d \leq \frac{\sqrt{2}}{2}r$; 2) $N_{\text{CS}}=5$, if $\frac{\sqrt{2}}{2}r < d \leq r$.

For multi-hop networks, the path for a data flow is found by a routing protocol. We only focus on the performance analysis of the MAC protocol here. To avoid discussing the routing protocol, we make some assumptions on flow configurations.
We assume that every node in the considered network can be a source node. For a node (e.g., node $n_{1}$ in Fig. \ref{figFlowConfiguration}), each of its neighbors (i.e., all nodes except for $n_{1}$ in the CS range of $n_{1}$) can be its next hop. We configure each node as the source node of $(N_{\text{CS}}-1)$ (i.e., the number of the node's neighbors) flows. For simplicity, we also assume that all flows have the same path
length $N_{\text{path}}$ and pass through nodes that are connected as a straight line, where the path length of a flow is defined as the number of nodes in the flow. Flows with $n_{1}$ as the source node are shown in Fig. \ref{figFlowConfiguration}, where $N_{\text{path}}=4$ in this example and one of these flows is from $n_{1}$ to $n_{4}$. This flow configuration is repeated for every node in the network, so that there can be multiple (partly overlapping) flows crossing each node.

To simplify our analysis, we also assume that only one node at a particular location of the flow can initiate an E2E-KIC process for this flow. This ideal assumption is consistent with one of the core ideas of our E2E-KIC MAC design, i.e., to avoid unnecessary contentions among nodes in a same flow. When seen from the source node to the destination node of a particular flow, we say that the source node is the first node in this flow, and the destination node is the last node in this flow. We use $n_{i}$ to denote the $i$th node of a flow, where we note that $n_{i}$ can be different nodes when considering different flows. The variable $i$ is constant in our analysis, so that an E2E-process is always initiated by the $i$th node of the flow (i.e., node $n_i$). In Fig. \ref{figAnE2EKICProcess} and the remaining discussion, we use $Y$ to refer to a particular (possibly arbitrary) node. When node $Y$ is an initiating node for the current flow under consideration, we call it $n_{i}$; otherwise, we directly call it $Y$. When node $n_{i}$ initiates an E2E-KIC process, the node $n_{l}$ $(l\neq i)$ involved in this E2E-KIC process is named as the $l$th cooperator. As shown in the example in Fig. \ref{figAnE2EKICProcess}, based on the above assumptions, the node $n_{i+1}$ can be the $l$th cooperator of $(N_{\text{CS}}-1)$ flows (which originate at nodes marked with grid background in the figure). It is because although there are multiple overlapped flows passing through $n_{i+1}$, only $(N_{\text{CS}}-1)$ of them have source nodes that are $l$ hops away from $n_{i+1}$.
%Little cycles with grid background in Fig. \ref{figAnE2EKICProcess} show the $(N_{\text{CS}}-1)$ source nodes under the condition that $l=2$.
%%%%%%%%%%%%%%%%%%%%%%%%%%%%%%%%%%%%%%%%%%%%%%%%%%%%%%%%%%%%%%%%%%%%%%%%%
%%%%%%%%%%%%%%%%%%%%%%%%%%%%%%%%%%%%%%%%%%%%%%%%%%%%%%%%%%%%%%%%%%%%%%%%%
\subsubsection{Calculation of Cooperation Probability}
The probability that a node is successfully cooperated as the $l$th cooperator is defined as $P_{\text{c}}(l)$. We can find the overall cooperation probability $P_{\text{c}}$ from $P_{\text{c}}(l)$ (see (\ref{AnotherRelation}) below). For convenience, let $P_{\text{Silent}}=1- P_{\text{t}} -P_{\text{c}}$ denote the probability that a node does not transmit in a backoff slot, i.e., it neither initiates E2E-KIC nor is being cooperated. Let $N_{\text{hid}}$ denote the number of hidden terminals seen from a particular node in a particular transmission, and $\delta$ denote the average time duration between two adjacent state transition events. The expression of $\delta$ will be given in Section \ref{DurationOfStateTxi}.

Because the node placements are symmetric, the initiating node $n_{i}$ can initiate one of $(N_{\text{CS}}-1)$ E2E-KIC processes with probability $P_{\text{t}}/(N_{\text{CS}}-1)$. As shown in Fig. \ref{figAnE2EKICProcess}, the next hop $n_{i+1}$ of $n_{i}$ can be successfully cooperated if: 1) neighbors of $n_{i+1}$ do not transmit in the same backoff slot; and 2) hidden terminals keep silent during the entire RTS transmission time (which is equal to the CTS transmission time). The probability of satisfying the first condition is $(P_{\text{Silent}})^{(N_{\text{CS}}-1)}$, where we recall that node events are assumed to be independent, as introduced earlier. The probability of satisfying the second condition is defined as $P_{\text{nhid}}$. Similarly to \cite{ThroughputTingChao}, we express $P_{\text{nhid}}$ as $P_{\text{nhid}}={\left( P_{\text{Silent}} \right)}^{{{N}_{\text{hid}}}\left( T_{\text{CTS}}/\delta \right)}$, which is intuitively the probability that all the ${N}_{\text{hid}}$ terminals keep silent during $T_{\text{CTS}}$. Because node $n_{i+1}$ can be the $(l=i+1)$th cooperator of $(N_{\text{CS}}-1)$ E2E-KIC processes, the probability that $n_{i+1}$ is successfully cooperated as the $(l=i+1)$th cooperator can be written as
\begin{equation}
P_{\text{c}}(l=i+1) = (N_{\text{CS}}-1) \cdot \left(\frac{P_{\text{t}}}{N_{\text{CS}}-1} \right)\cdot \left( P_{\text{Silent}}\right)^{(N_{\text{CS}}-1)}P_{\text{nhid}}\,.
\end{equation}

%As shown in Fig. \ref{figAnE2EKICProcess}, RTS transmission from $n_{i}$ to $n_{i+1}$ is successful when neighbors of $n_{i+1}$ (nodes within the gray disk-shaped region) keep silent during the RTS transmission,
If $n_{i+1}$ is successfully cooperated, $n_{i+1}$ has successfully received RTS from $n_{i}$ and all the neighbors of $n_{i}$ will keep silent during the remaining time of this E2E-KIC process. Node $n_{i+1}$ responses with a CTS frame in this case. The neighbors of $n_{i+1}$ can successfully receive this CTS frame with probability $\left( P_{\text{Silent}} \right)^{N_{\text{hid}}}$. The CTS sent by $n_{i+1}$ also serves for the purpose of requesting cooperation with its next hop $n_{i+2}$, which will only be successful if the hidden terminals (labeled by the text in Fig. \ref{figAnE2EKICProcess}) among the CS range of $n_{i+2}$ keep silent during the time when $n_{i+1}$ transmits the CTS frame. Therefore, the probability that node $n_{i+2}$ is cooperated as the $(l=i+2)$th cooperator can be written as
\begin{equation}
P_{\text{c}}(l=i+2) = P_{\text{c}}(l=i+1) \left( P_{\text{Silent}} \right)^{N_{\text{hid}}}P_{\text{nhid}}\,.
\end{equation}

Repeating the above steps, a node is successfully cooperated as the $l$th $(l\in [1,N_{\text{path}}],\ l \neq i)$ node with probability
\begin{IEEEeqnarray}{lll}
P_{\text{c}}\left( l \right) &{}={}& P_{\text{t}} \left( P_{\text{Silent}} \right)^{(n_{\text{CS}}-1)} \left( P_{\text{Silent}}\right)^{(\left| l -i\right| - 1) n_{\text{hid}}} \nonumber \\
{}&{}&\left( P_{\text{nhid}} \right)^{\left| l-i \right|+1-\mathbf{1}_{l-i+1}}\,,
\end{IEEEeqnarray}
where $\mathbf{1}_{x}=1$, if $x>0$; otherwise, $\mathbf{1}_{x}=0$. The term $\mathbf{1}_{l-i+1}$ is because $n_{i-1}$ waits for one CTS duration before it transmits its own CTS, as shown in Fig. \ref{timingdiagram}. The overall cooperation probability of a node can be expressed as
\begin{equation}
P_{\text{c}}=\sum_{l=1,l\neq i}^{N_{\text{path}}}P_{\text{c}}\left( l \right)\,,
\label{AnotherRelation}
\end{equation}
which gives another relationship between $P_{\text{t}}$ and $P_{\text{c}}$.
\subsection{Average Duration between Adjacent State Transitions}\label{DurationOfStateTxi}
We note that (\ref{AnotherRelation}) contains a parameter $\delta$, which is embedded in $P_{\text{nhid}}$. We determine $\delta$ in the following by discussing the events that lead to state transitions. Due to the symmetry of node placements, our analysis is based on the viewpoint of a single node.

Let node $Y$ be the considered node. Similarly to \cite{ThroughputTingChao}, the possible events that lead to state transition of $Y$ are: 1) when keeping silent, $Y$ countdowns its backoff counter if the channel is idle or at least one of its neighbors ends its transmission; 2) when initiating an E2E-KIC process or being cooperated, $Y$ resets its backoff counter. The events can be further classified into five cases. Let $\delta_{j}$ denote the average duration of case~$j$ multiplied by the probability that case~$j$ occurs. We analyze each case in the following, and the results are summarized in Table \ref{AllTable}. The value of $\delta$ can be evaluated by
\begin{equation}
\delta = \sum_{j=1}^{5}\left(\delta_{j}\right)\,.
\label{eq:deltaOverallExpression}
\end{equation}

\begin{table*}[!t]
	\centering
	\caption{Five cases leading to state transitions}
	\begin{tabular}{c||c||c||c}
		\hline\hline
		\multirow{1}*{Case $j$} & Probability & Duration of sub-case & $\delta_j$ \\
		\hline\hline
		\multirow{1}*{Case 1} & $P_{\text{Silent}}P_{X\text{Tx}\_1}$ & $T_{X\text{Tx}\_1}$ & $\delta_{1}=P_{\text{Silent}}\cdot P_{X\text{Tx}\_1}\cdot T_{X\text{Tx}\_1}$\\
		\hline\hline
		\multirow{1}*{Case 2} & $P_{\text{Silent}}P_{X\text{Tx}\_2}(l)$ & $T_{X\text{Tx}\_2}(l)$ & $\delta_{2}=\sum_{l=1,l\neq i}^{N_{\text{path}}}{P_{\text{Silent}}\cdot (P_{X\text{Tx}\_2}(l)\cdot T_{X\text{Tx}\_2}(l))}$\\
		\hline\hline
		\multirow{1}*{Case 3} & $P_{\text{Silent}}P_{X\text{Tx}\_3}$ & $T_{X\text{Tx}\_3}$ & $\delta_{3}=P_{\text{Silent}}\cdot P_{X\text{Tx}\_3}\cdot T_{X\text{Tx}\_3}$\\
		\hline\hline
		\multirow{3}*{Case 4} & $P_{n_{i}\text{Tx}\_1}$ & $T_{n_{i}\text{Tx}\_1}$ &\\\cline{2-3}
		& $P_{n_{i}\text{Tx}\_2}$ & $T_{n_{i}\text{Tx}\_2}$ & \\\cline{2-3}
		& $P_{n_{i}\text{Tx}\_3}$ & $T_{n_{i}\text{Tx}\_3}$ & \multirow{-3}*{$\delta_{4}=\sum_{q=1}^{3}{(P_{n_{i}\text{Tx}\_q}\cdot T_{n_{i}\text{Tx}\_q})}$}\\
		\hline\hline
		\multirow{1}*{Case 5} & $P_{Y\text{Tx}\_5}(l)$ & $T_{Y\text{Tx}\_5}(l)$ & $\delta_{5}=\sum_{l=1,l\neq i}^{N_{\text{path}}}{(P_{Y\text{Tx}\_5}(l)\cdot T_{Y\text{Tx}\_5}(l))}$\\
		\hline\hline
	\end{tabular}
	\label{AllTable}
\end{table*}

\subsubsection{The Considered Node Keeps Silent (Cases 1--3)}
When $Y$ keeps silent, we have three cases: 1) only one neighbor of $Y$ initiates an E2E-KIC process, 2) only one neighbor of $Y$ is cooperated, and 3) more than one neighbor of $Y$ initiate E2E-KIC processes or are cooperated.

We first consider Case 1. Generally, a particular neighbor of $Y$ can arbitrarily initiate one of $(N_{\text{CS}}-1)$ E2E-KIC processes (flows). However, the initiated E2E-KIC flow cannot involve $Y$, because of the condition that $Y$ is idle. Let $N_{\text{init}}~(N_{\text{init}}<(N_{\text{CS}}-1))$ denote the number of E2E-KIC flows that can be initiated by a neighbor of $Y$ in this case. For a neighbor of $Y$, the initiation probability is $(N_{\text{init}}/{(N_{\text{CS}}-1)})P_{\text{t}}$. Therefore, noting that $Y$ is idle, the probability that only one neighbor of $Y$ initiates an E2E-KIC process can be written as
\begin{IEEEeqnarray}{ll}
\label{Xtrans1}
P_{X\text{Tx}\_1}&{}=(N_{\text{CS}}-1) \cdot \frac{N_{\text{init}}}{N_{\text{CS}}-1} P_{\text{t}}\cdot\left( P_{\text{Silent}} 
 \right)^{(N_{\text{CS}}-2)} \nonumber \\
&{}=N_{\text{init}}P_{\text{t}}\left( P_{\text{Silent}} \right)^{(N_{\text{CS}}-2)}\,.
\end{IEEEeqnarray}
Equation (\ref{Xtrans1}) can be explained as that one of $Y$'s $(N_{\text{CS}}-1)$ neighbors initiates an E2E-KIC process with probability $(N_{\text{init}}/{(N_{\text{CS}}-1)})P_{\text{t}}$, and other neighbors of $Y$ keep silent. In this case, node $Y$ can successfully receive this RTS frame and stays in the idle state for the time specified by the NAV field of the RTS frame. Unlike that in Section \ref{sectMAC}, the RTS NAV in this section is set to the time of a whole E2E-KIC process to simplify our expressions, which can be written as
\begin{IEEEeqnarray}{ll}
T^{'}_{\text{NAV}}(\text{RTS})={}&T_{\text{RTS}}+(\max(A+1,P))(T_{\text{SIFS}}+T_{\text{CTS}}) \nonumber \\
&+T_{\text{DATA}}+2T_{\text{frame-diff}}+T_\textrm{before-ACK} \nonumber \\
&+T_{\text{SIFS}}+T_{\text{ACK}}+T_{\text{DIFS}}.
\end{IEEEeqnarray}
Therefore, the duration of this case is $T_{X\text{Tx}\_1}=T^{'}_{\text{NAV}}(\text{RTS})$, and we have $\delta_{1}=P_{\text{Silent}}\cdot P_{X\text{Tx}\_1}\cdot T_{X\text{Tx}\_1}$.

Then, we consider the Case 2 that only one neighbor of $Y$ is cooperated. According to a similar analysis to that of Case 1, one of the $Y$'s neighbors (defined as $X$) in Case 2 is cooperated as the $l$th node with probability
\begin{IEEEeqnarray}{ll}
\label{Xtrans2}
P_{X\text{Tx}\_2}(l) ={}&(N_{\text{CS}}-1) \cdot \frac{N_{\text{Init}}}{N_{\text{CS}}-1}P_{\text{c}}(l) \cdot \left( P_{\text{Silent}} \right)^{(N_{\text{CS}}-2)} \nonumber \\
{}&=N_{\text{Init}}P_{\text{c}}(l)\left( P_{\text{Silent}} \right)^{(N_{\text{CS}}-2)}\,.
\end{IEEEeqnarray}
After being successfully cooperated, node $X$ responds a CTS frame, which can be received by $Y$ because all other nodes among the CS range of $Y$ keep silent. Node $Y$ stays idle for the time specified by the NAV field of the CTS frame, which is $T^{'}_{\text{NAV}}(\text{CTS},n_{l})=T^{'}_{\text{NAV}}(\text{RTS})$. The duration of this case is $T_{X\text{Tx}\_2}(l)=T^{'}_{\text{NAV}}(\text{CTS},n_{l})$, and we have $\delta_{2}=\sum_{l=1,l\neq i}^{N_{\text{path}}}P_{\text{Silent}}\cdot (P_{X\text{Tx}\_2}(l)\cdot T_{X\text{Tx}\_2}(l))$.

Finally, we consider Case 3 that more than one neighbors of $Y$ initiate transmissions or are cooperated.
The corresponding probability of this case is calculated by
$P_{X\text{Tx}\_3}=P_{X\text{Tx}\_\text{all}}-\left(P_{X\text{Tx}\_1}+\sum_{l}P_{X\text{Tx}\_2}(l)\right)$, where $P_{X\text{Tx}\_\text{all}}=1 - \left( P_{\text{Silent}} \right) ^{N_{\text{CS}}-1}$ is the probability that under the condition that $Y$ is idle, at least one neighbor of node $Y$ initiates transmission or is cooperated.

There are too many sub-cases in this case, so it is hard to discuss them one by one. For simplicity, we only consider the special case that two neighbors of $Y$ respectively initiate two E2E-KIC processes at the same time, which is a typical sub-case in Case 3. For each flow, after successfully receiving the RTS frame, the initiating node's next hop sends a CTS frame. If any initiating node has successfully received a CTS frame from its next hop\footnote{We ignore the low-probability event that collision occurs at the two initiating nodes when they are receiving their corresponding CTS frames.}, the two initiating nodes are assumed to occupy the channel for the time of a whole E2E-KIC process; otherwise, the channel is only occupied for the RTS transmission time. Therefore, the duration of this case is  $T_{X\text{Tx}\_3} \approx  (1-(1-P_{\text{nhid}})^2)T^{'}_{\text{NAV}}(\text{RTS})+(1-P_{\text{nhid}})^2(T_{\text{RTS}}+T_{\text{DIFS}})$, and $\delta_{3}=P_{\text{Silent}}\cdot P_{X\text{Tx}\_3}\cdot T_{X\text{Tx}\_3}$.
\subsubsection{Node $Y$ Initiates a Transmission (Case 4)}
In this case, $Y$ is an initiating node, so we call it $n_{i}$ for convenience. When $n_{i}$ initiates an E2E-KIC process, it sends an RTS frame to its neighbors $n_{i-1}$ and $n_{i+1}$ (see Figs. \ref{timingdiagram} and \ref{figAnE2EKICProcess}). We have three sub-cases: 1) nodes\footnote{The terms ``nodes'' in the three sub-cases mean ``nodes except for $n_{i}$''.} within the CS range of $n_{i-1}$ and $n_{i+1}$ do not transmit; 2) nodes within the CS range of $n_{i-1}$ (or $n_{i+1}$) do not transmit, while at least one node within the CS range of $n_{i+1}$ (or $n_{i-1}$) transmits; 3) nodes within the CS range of $n_{i-1}$ and $n_{i+1}$ transmit. The probabilities of these sub-cases  are respectively: 1) $P_{n_{i}\text{Tx}\_1}=P_{\text{t}} P_{\text{CS}\_\text{Silent}}^2$, 2) $P_{n_{i}\text{Tx}\_2}=2P_{\text{t}} P_{\text{CS}\_\text{Silent}}\left(1-P_{\text{CS}\_\text{Silent}}\right)$, and 3) $P_{n_{i}\text{Tx}\_3}=P_{\text{t}} \left(1-P_{\text{CS}\_\text{ Silent}}\right)^2$, where $P_{\text{CS}\_\text{Silent}}=\left( P_{\text{Silent}} \right) ^{N_{\text{CS}}-1}$ denotes the probability that except for a particular node, nodes within the CS range do not transmit in the current backoff slot.

Considering the first sub-case, if neither of nodes $n_{i-1}$ and $n_{i+1}$ successfully receive the RTS frame, the duration is equal to the CTS timeout time; otherwise, the duration is the time of an entire E2E-KIC process. The duration of the second sub-case can be derived through a similar analysis. The duration of the third sub-case is always the CTS timeout time, because both nodes $n_{i-1}$ and $n_{i+1}$ cannot successfully receive the RTS frame due to collisions. Overall, the corresponding duration of the first case is:
\begin{IEEEeqnarray}{ll}
T_{n_{i}\text{Tx}\_1}&{}=\left(1-P_{\text{nhid}}\right)^2 T_{\text{CTS-timeOut}} \nonumber \\
&{}+ \left(1-\left(1-P_{\text{nhid}}\right)^2\right)T^{'}_{\text{NAV}}(\text{RTS})\,, 
\end{IEEEeqnarray}
where $T_{\text{CTS-timeOut}}=T_{\text{RTS}}+T_{\text{SIFS}}+T_{\text{CTS}}+T_{\text{DIFS}}$ denotes the CTS timeout time. The durations of the second and third cases are:
\begin{IEEEeqnarray}{ll}
T_{n_{i}\text{Tx}\_2}=P_{\text{nhid}}T^{'}_{\text{NAV}}(\text{RTS})+\left(1-P_{\text{nhid}}\right)T_{\text{CTS-timeOut}}\,. \\
T_{n_{i}\text{Tx}\_3}=T_{\text{CTS-timeOut}}\,.
\end{IEEEeqnarray}
Then,  $\delta_{4}=\sum_{q=1}^{3}{(P_{n_{i}\text{Tx}\_q}\cdot T_{n_{i}\text{Tx}\_q})}$.
\subsubsection{Node $Y$ is Cooperated (Case 5)}

In Case 5, node $Y$ is cooperated as the $l$th node in an E2E-KIC process, and the corresponding probability is $P_{Y\text{Tx}\_5}(l)=P_{\text{c}}(l)$. The corresponding duration is $T_{Y\text{Tx}\_5}(l)=T^{'}_{\text{NAV}}(\text{CTS},n_{l})$, and $\delta_{5}=\sum_{l=1,l\neq i}^{N_{\text{path}}}{(P_{Y\text{Tx}\_5}(l)\cdot T_{Y\text{Tx}\_5}(l))}$.

\subsection{Throughput Formulation}\label{ThrouFormu}
%The throughput is equal to the ratio of the average successfully transmitted payload and the average duration. If an anterior node $n_{k}$ $(k \in [1,i-1])$ is cooperated by its next hop, it can successfully send its data frame with a high probability, because neighbors of its next hop have sensed that the channel is busy. After being cooperated by the previous hop, a posterior node $n_{j}$ $(j \in [i+1,N_{\text{path}}-1])$ can successfully send its data frame if: 1) node $n_{j+1}$ (i.e., the next hop of $n_{j}$) is successfully cooperated, 2) and the hidden nodes of $n_{j+1}$ is informed the channel occupation time by $n_{j}$. The condition 2 can be satisfied if these hidden terminals do not contend channel with $n_{j+1}$ and keep silent when $n_{j+1}$ responses the requisition from $n_{j}$. Therefore, both conditions have the same probability $(P_{\text{Silent}})^{N_{\text{hid}}}(P_{\text{nhid}})$. Let $L$ be the length of the payload. Let $P_{\text{c}\_\text{an}}=\sum_{k=1}^{i-1}P_{\text{c}}(k)$ (or $P_{\text{c}\_\text{pos}}=\sum_{j=i+1}^{N_{\text{path}}-1}P_{\text{c}}(j)$) denote the probability that a node is cooperated as an anterior (or a posterior) node. The per-node throughput can be formulated as
%\begin{equation}
%S_{1}=\frac{\left(P_{\text{c}\_\text{an}}+\left(P_{\text{c}\_\text{pos}}+P_{\text{t}}\right)(P_{\text{Silent}})^{n_{\text{CS}}-1}(P_{\text{nhid}})^2\right) L}{\delta}\,.
%\label{Throughput1}
%\end{equation}
The throughput is defined as the successfully transmitted payload (data) per node per unit time. It can be expressed as the ratio of the amount of successfully transmitted data to the average duration between two adjacent state transitions.
%With E2E-KIC, the throughput of a node (e.g., node $Y$ in Fig. \ref{figAnE2EKICProcess}) can be obtained through successfully transmitting data when the node either initiates a transmission or is cooperated.
We note that from (\ref{TimeDomiEqn}), (\ref{AnotherRelation}), and (\ref{eq:deltaOverallExpression}), we can numerically solve for the variables  $P_{\text{t}}$, $P_{\text{c}}$, and $\delta$. Therefore, we express the throughput with these three variables (and other variables that are functions of them) in the following discussion.

An arbitrary node $Y$ successfully transmits data to another node when the following two conditions are \emph{both} satisfied: 1) its next hop is successfully cooperated by node $Y$, 2) and its second hop is successfully cooperated by its next hop. It is because that as shown in Fig. \ref{figAnE2EKICProcess}, if the next and second hops are both successfully cooperated, the neighbors of the next hop (including hidden terminals) can be informed of the channel-occupation time and keep silent during the time of data exchange\footnote{We ignore the low-probability events that collisions occur at neighbors of $n_{i+1}$ as nodes $Y$ and $n_{i+1}$ are waiting for their intended CTS frames.}. When node $Y$ is an initiating node with initiation probability $P_{\text{t}}$, the probabilities of satisfying the above two conditions are respectively $(P_{\text{Silent}})^{N_{\text{CS}}-1}(P_{\text{nhid}})$ and $(P_{\text{Silent}})^{N_{\text{nid}}}(P_{\text{nhid}})$. Let $L$ be the size of the transmitted data in each channel access. The average amount of successfully transmitted data by an initiating node can be expressed as ${P_{\text{t}}(P_{\text{Silent}})^{N_{\text{CS}}-1}(P_{\text{Silent}})^{N_{\text{nid}}}(P_{\text{nhid}})^2L}$.  When node $Y$ is a cooperator with cooperation probability $P_{\text{c}}$, the probabilities of satisfying either Condition 1 or Condition 2 above are the same, and they are equal to $(P_{\text{Silent}})^{N_{\text{hid}}}(P_{\text{nhid}})$, because in this case, collisions can be caused only by the hidden terminals as discussed in Section \ref{sec:coopProba}. The average amount successfully transmitted data by a cooperator can be written as ${P_{\text{c}}((P_{\text{Silent}})^{N_{\text{nid}}}P_{\text{nhid}})^2L}$. Therefore, the saturation throughput can be expressed as
\begin{IEEEeqnarray}{lll}
S&{}={}&\frac{P_{\text{t}}(P_{\text{Silent}})^{N_{\text{CS}}-1}(P_{\text{Silent}})^{N_{\text{nid}}}(P_{\text{nhid}})^2L}{\delta}\nonumber \\
{}&{}&+\frac{P_{\text{c}}((P_{\text{Silent}})^{N_{\text{nid}}}P_{\text{nhid}})^2L}{\delta}\,.
\label{EqnThroughput}
\end{IEEEeqnarray}
From (\ref{EqnThroughput}), we can see that avoiding unnecessary contentions from hidden terminals (thereby increasing $P_{\text{nhid}}$) increases the amount of successfully transmitted data, thus increases the throughput. Meanwhile, limiting control overhead decreases the average duration between two adjacent state transitions, which also increases the throughput. This is in accordance with the design principles of E2E-KIC MAC presented in Section \ref{sub:basicPrinciple}.
\section{Simulation Results} \label{sectSim}
The performance of E2E-KIC is evaluated with our discrete-event simulator jointly developed with MATLAB and C \cite[Appendix A]{MAC}. We first apply the E2E-KIC scheme into a two-dimensional grid topology to confirm our theoretical performance analysis. Then, we simulate the E2E-KIC scheme in some other typical multi-hop topologies to investigate its performance in general cases.
\subsection{Performance in Two-Dimensional Grid Topology} \label{sectSim:gridTopo}
We consider a two-dimensional grid topology with $10\times 10$ nodes. The transmission radius $r$ is set to $200$~m. The node distance $d$ is set to $110$ m or $150$ m to cover the two topology cases: $N_{\text{CS}}=9$ and $N_{\text{CS}}=5$. The path size $N_{\text{path}}$ is set to $10$, and both hop limits $N_{\text{an}}$ and $N_{\text{pos}}$ are set to 5. Other parameter settings can be found in \cite{OurTechRep}. For an E2E-KIC process, the initiating node is always the sixth node of the corresponding flow path. To evaluate the overall performance, each simulation is run for $10$~s with $10$ different random seeds. Contention window size is fixed in each simulation.

\begin{figure}[!t]
	\centerline{\subfigure[]{\includegraphics[width=0.24\textwidth]{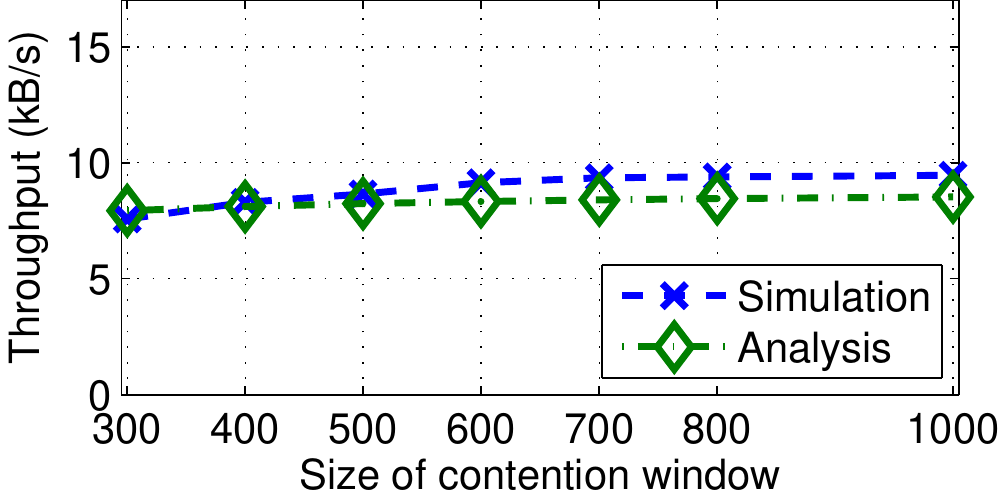}
			\label{RegularGridDistance110}}
		\hfil
		\subfigure[]{\includegraphics[width=0.24\textwidth]{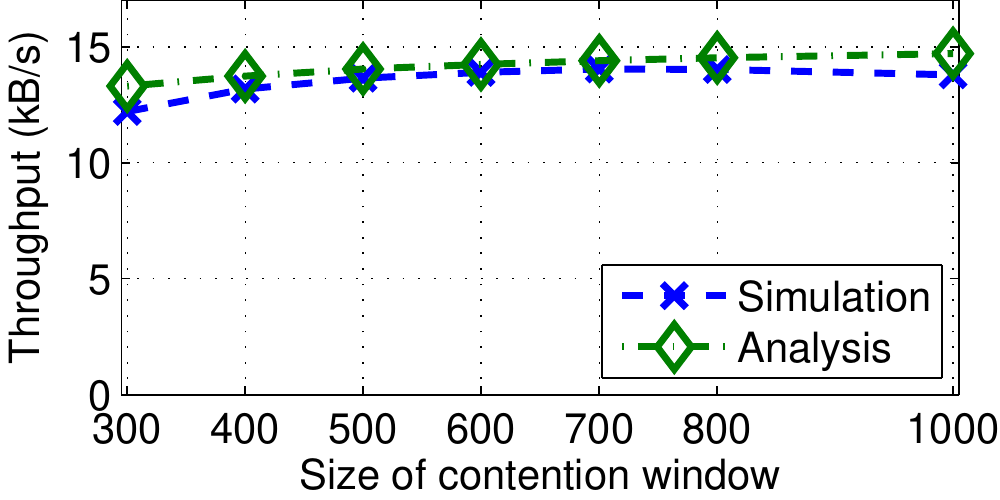}
			\label{RegularGridDistance150}}}
	\caption{Performances vs. size of contention window in two-dimensional grid network: (a) $d=110$, (b) $d=150$.}
	\label{fig_RegularGrid}
	\vspace*{-2pt}
\end{figure}

Fig. \ref{fig_RegularGrid} shows the saturation throughput under different contention window sizes. The simulation results are calculated while filtering out the throughput of nodes located at the edges of the considered topology to reduce boundary effects. As shown in Fig. \ref{fig_RegularGrid}, the analytical results of both topology cases are very consistent with the simulation results. The differences between them are caused by several assumptions/simplifications we have made. The dense topology (where $N_{\text{CS}}=9$) achieves a smaller throughput compared to the sparse topology (where $N_{\text{CS}}=5$), due to intense channel contentions. In both topologies, the saturation throughput increases with the increase of the contention window size if the contention-window size is smaller than $700$, because a larger contention window alleviates collisions. Afterwards, the saturation throughput stays similar (with a slight increase or decrease), which corresponds to the maximal saturation throughput of the considered scenario.
\subsection{Performance in other Multi-hop Topologies}
In this subsection, we consider general scenarios and realistic channel modeling following \cite[Appendix A]{MAC}. The transmission power is set to $0$~dBm ($1$~mW). The channel power gain is $1/D_{i,j}^4$, where $D_{i,j}$ is the distance between nodes $n_{i}$ and $n_{j}$ in meters. A node can \emph{receive} a signal from the distance of $200$~m. The background noise density is $-174$~dBm/Hz with $6$~dB noise figure. The receiver's clear channel assessment (CCA) sensitivity is set to $-106$~dBm. With these settings, a node can \emph{sense} signal transmissions from approximately $450$~m. The additional waiting time $T_{\text{wait}}$ is $0.05$~s. Contention window size is calculated with the algorithm described in Section \ref{sec:scheAlgo}.

The performance of E2E-KIC is compared with other approaches (including both physical-layer techniques and their MAC protocols) that support one-hop or two-hop KIC, and also with the PR method represented by the IEEE 802.11 MAC protocol. Considering both PNC and FD, two KIC-supported MAC protocols are selected: 1) the overlapped-carrier sensing medium access (OCSMA) MAC protocol in \cite{OCSMA}, i.e., the PNC-supported MAC protocol; 2) the EF-FD MAC protocol in \cite{EFFD}. EF-FD MAC is considered for comparison in our simulations because it supports FD in general multi-hop networks and includes an algorithm to avoid unnecessary contentions. It also eliminates the problem of hidden node with the combination of FD and busy tone. To evaluate the overall performance, each simulation is run for a duration of $50$~s and with $10$ different random seeds.
\subsubsection{Performance under Different Packet Rates in Chain Topology}
We first consider a chain topology with $7$ nodes. The distances between adjacent nodes are fixed at $200$~m. One flow from the first node to the end node is configured.

\begin{figure}[!t]
\centerline{\subfigure[]{\includegraphics[width=0.37\textwidth]{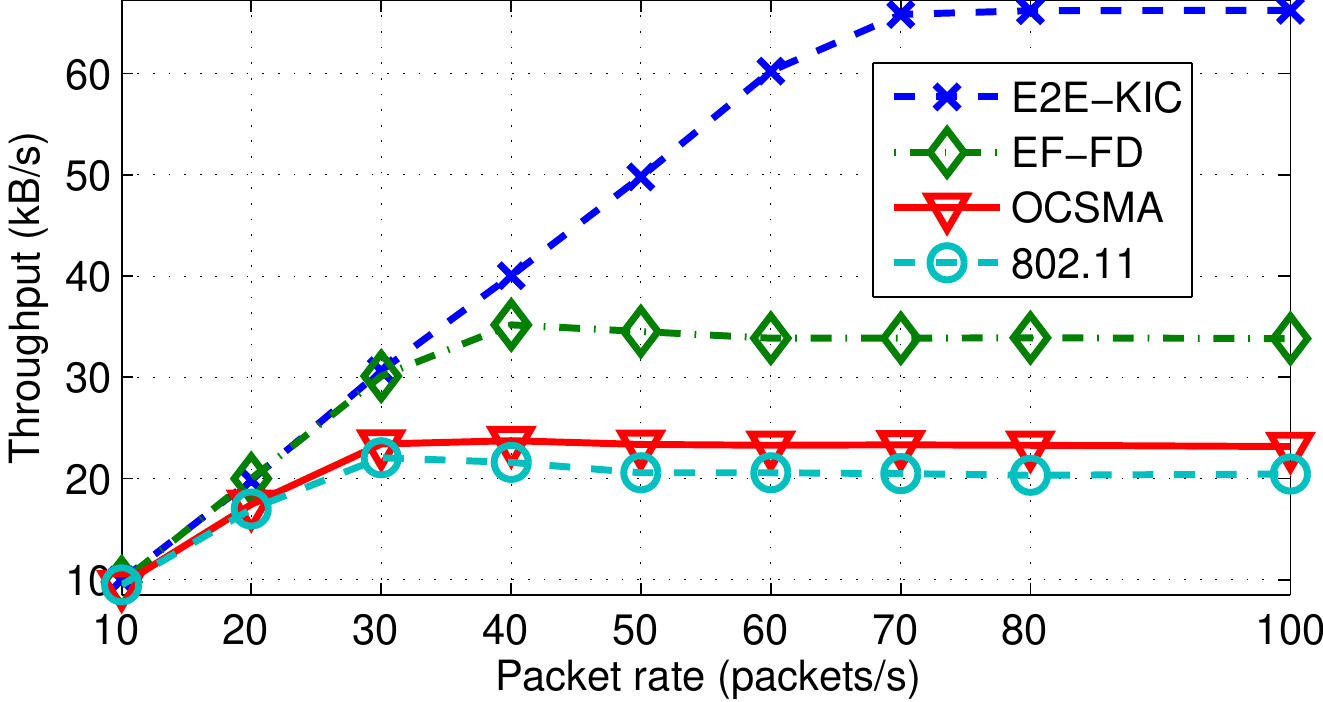}
\label{LinearPktRatethroughput}}}
\centerline{\subfigure[]{\includegraphics[width=0.37\textwidth]{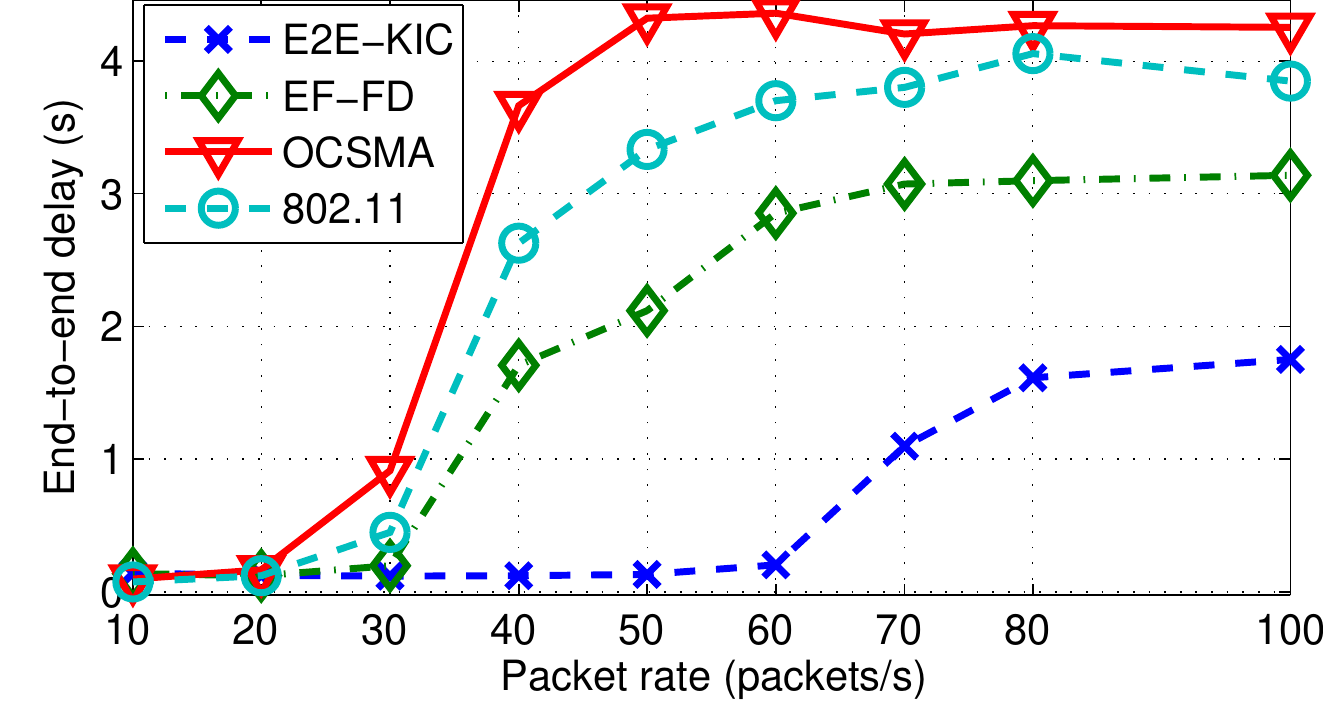}
\label{LinearPktRatedelay}}}
\caption{Performances vs. packet rate in a chain topology: (a) Throughput, (b) End-to-end delay.}
\label{fig_sim1}
\vspace*{-2pt}
\end{figure}
%
%\begin{figure}[!t]
%\centering \includegraphics[width=0.4\textwidth]{picture/averageD-linear-PacketRate.pdf}
%\caption{End-to-end delay vs. packet rate in the linear topology.}
%\label{LinearPktRatedelay}
%\end{figure}
The throughput under various packet rates in the chain topology is shown in Fig.~\ref{LinearPktRatethroughput}. We can observe that the different schemes have similar throughput performances when the packet rate is smaller than $20$~packets/s. The reason is that when the traffic load is low, the wireless channel is idle for most of the time. Even with the PR method, the packet can be forwarded to the next node within a short time after receiving it, so the benefit of E2E-KIC is not obvious. The highest throughputs of EF-FD, OCSMA, and IEEE 802.11 are achieved at about $40$~packets/s, while the highest throughput of E2E-KIC is achieved at about $70$~packets/s. It is obvious that the proposed E2E-KIC mechanism achieves the highest throughput. Specifically, the throughput gains of E2E-KIC over EF-FD and IEEE 802.11 are close to $2$ and $3$ respectively, which are in accord with our throughput analysis in Section \ref{TAnalysis}. OCSMA achieves the lowest throughput among the KIC-supported schemes, because it lacks a method of reducing unnecessary contentions.

The end-to-end delays under different packet rates are shown in Fig.~\ref{LinearPktRatedelay}. We can observe that the delay of E2E-KIC is lowest, because more nodes concurrently transmit and the MAC protocol also reduces unnecessary channel contentions. We can also find that the delay of EF-FD is lower than that of IEEE 802.11. OCSMA has the highest delay, because it does not consider the exceptional case that a node is informed to perform PNC but has no packet to send, which increases the packet exchange time.
\subsubsection{Performance under Different Numbers of Nodes in Chain Topology}
Now, we study the performance under different numbers of nodes in the chain topology. The distance between two adjacent nodes is still $200$~m, and one flow from the first node to the last node is configured. The packet rate is fixed at $100$~packets/s.

\begin{figure}[!t]
\centerline{\subfigure[]{\includegraphics[width=0.3\textwidth]{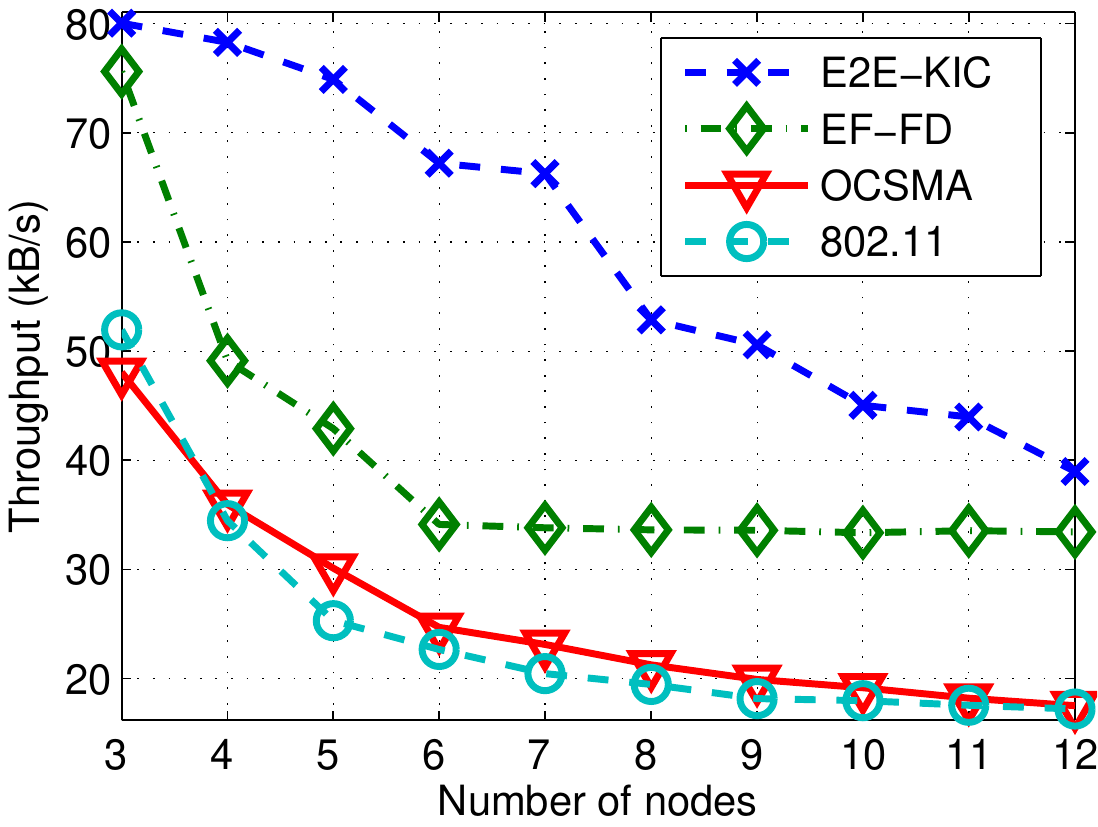}
\label{TLinearTopNodeNum}}}
\centerline{\subfigure[]{\includegraphics[width=0.3\textwidth]{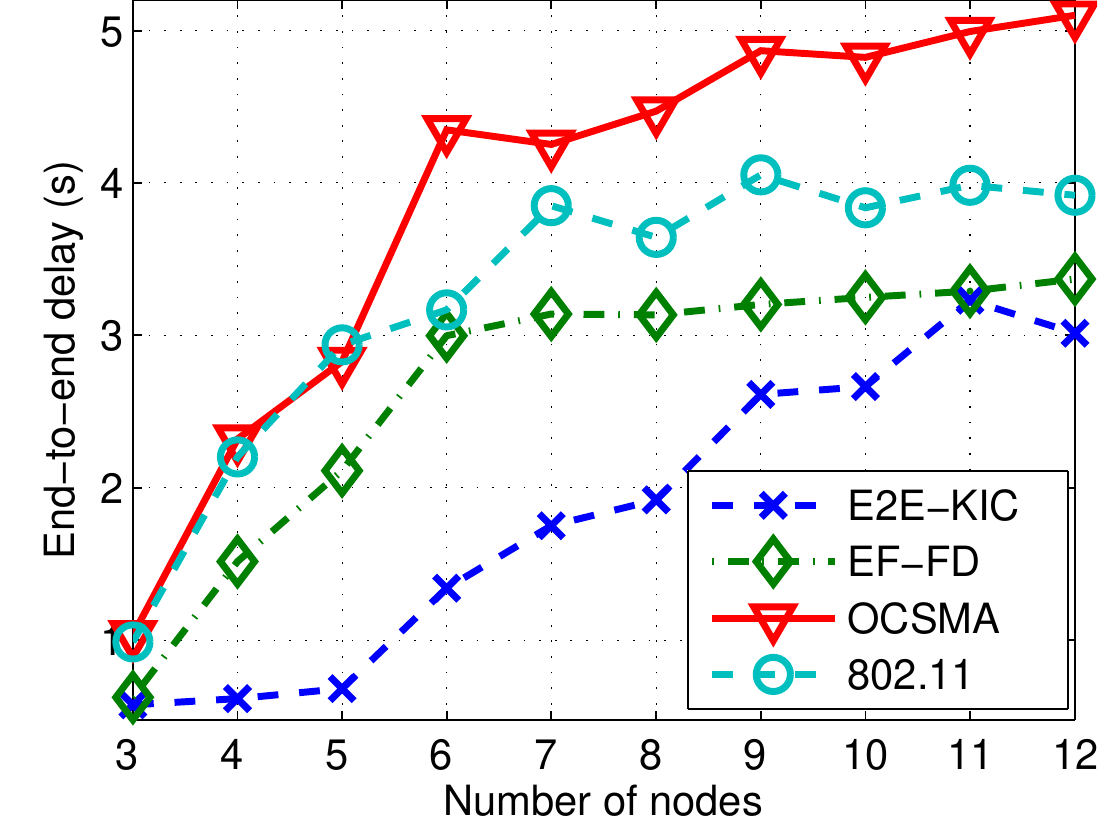}
\label{DLinearTopNodeNum}}}
\caption{Performances under different numbers of nodes in chain topologies: (a) Throughput, (b) End-to-end delay.}
\label{fig_sim2}
\vspace{-2pt}
\end{figure}

%
%
%\begin{figure}[!t]
%\centering \includegraphics[width=0.4\textwidth]{picture/AverageTLinearTopologyNodeNum.pdf}
%\caption{Throughput vs. node number in linear topologies.}
%\label{TLinearTopNodeNum}
%\end{figure}
Fig. \ref{TLinearTopNodeNum} shows the throughput under different numbers of nodes. The E2E-KIC MAC scheme outperforms the other three schemes in comparison. The maximum throughput gain of E2E-KIC over EF-FD is $1.97$ and the average throughput gain is $1.38$. We can also see that OCSMA has the lowest throughput among the KIC-supported schemes, for the same reasons as mentioned in the previous subsection. When increasing the number of nodes, the throughput of E2E-KIC decreases, because of the increasing overhead of control frames caused by increasing number of nodes. The same trend can be observed in the schemes under comparison.
%When the number of nodes is smaller than $6$, the throughputs of EF-FD, OCSMA, and IEEE 802.11 decrease with increasing number of nodes. After that, when further increasing the number of nodes, their throughputs slightly increase due to more opportunities of spatial reuse.

Fig.~\ref{DLinearTopNodeNum} shows the end-to-end delays under different number of nodes. We can observe that the end-to-end delay of E2E-KIC is lower than those of other protocols. OCSMA has the highest delay due to the aforementioned reasons.
%
%\begin{figure}[!t]
%\centering \includegraphics[width=0.4\textwidth]{picture/AverageDLinearTopologyNodeNum.pdf}
%\caption{End-to-end delay vs. node number in linear topologies.}
%\label{DLinearTopNodeNum}
%\end{figure}
%%
\subsubsection{Performance under Different Packet Rates in Random Topology}\label{RandomTopologySim}
Next, we consider a random topology with $20$ nodes distributed in a $800\times 800$~$\text{m}^2$ square, where $7$ flows with randomly selected source and destination nodes are configured.
\begin{figure}
\centerline{\subfigure[]{\includegraphics[width=0.3\textwidth]{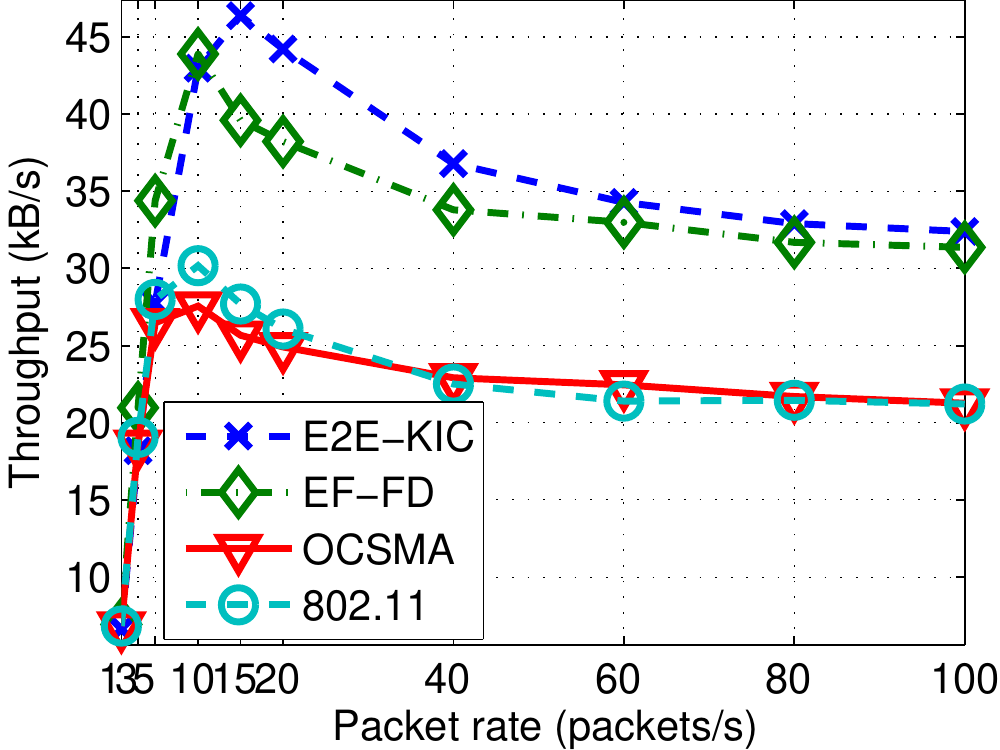}
\label{randomT}}}
\centerline{\subfigure[]{\includegraphics[width=0.3\textwidth]{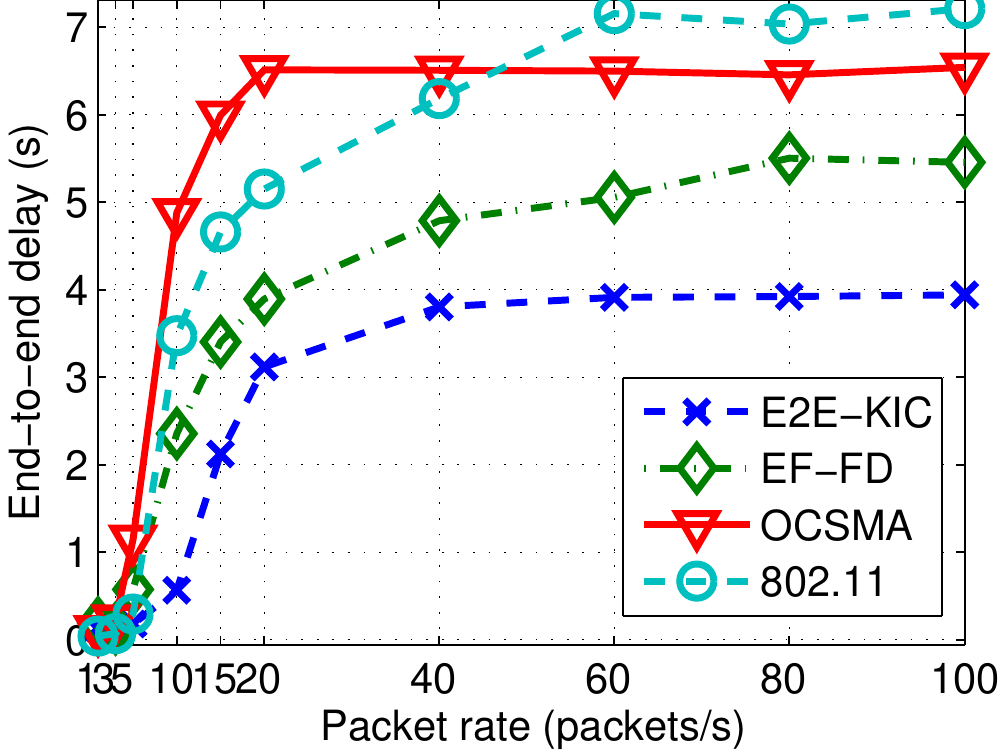}
\label{randomD}}}
\caption{Performances under different packet rates in random topologies: (a) Throughput, (b) End-to-end delay.}
\label{fig_sim3}
\vspace{-2pt}
\end{figure}
The average end-to-end throughput of the random topology under different packet rates is shown in Fig.~\ref{randomT}. It can be seen that the highest throughput of E2E-KIC is achieved at $15$~packets/s and slightly decreases with the packet rate afterwards because of intense channel contentions. For the same reason, other schemes behave similarly. We can also observe that E2E-KIC achieves the highest throughput among the evaluated schemes. Particularly, the maximum throughput gain of E2E-KIC over EF-FD is $1.30$ and the average throughput gain is $1.17$. The average throughput gains of E2E-KIC over IEEE 802.11 and OCSMA are $1.49$ and $1.66$, respectively.
%\begin{figure}
%\centering \includegraphics[width=0.4\textwidth]{picture/averageTrandom.pdf}
%\caption{Throughput vs. packet rate in the random topology.}
%\label{randomT}
%\end{figure}

The end-to-end delays under different packet rates are shown in Fig. \ref{randomD}. We can observe similar phenomena as in Fig. \ref{LinearPktRatedelay}. IEEE 802.11 MAC has the highest end-to-end delay when the packet rate is greater than $45$~packets/s. We note that when the packet rate increases, there are more KIC opportunities. By utilizing simultaneous transmissions, the KIC-supported schemes reduce the end-to-end delay compared with IEEE 802.11.
%\begin{figure}
%\centering \includegraphics[width=0.4\textwidth]{picture/averageDrandom.pdf}
%\caption{End-to-end delay vs. packet rate in the random topology.}
%\label{randomD}
%\end{figure}
\section{Conclusions and Future Work} \label{sectCon}
In this paper, we have proposed an E2E-KIC mechanism for general wireless multi-hop networks. The idea is that superposed signals can be successfully decoded by properly scheduling the transmission of each node belonging to the same data flow. Theoretical analysis shows that E2E-KIC has the potential to achieve the single-hop throughput performance in multi-hop transmissions. To effectively support E2E-KIC, we have also proposed the E2E-KIC MAC protocol. E2E-KIC MAC includes a new access-control mechanism, which maintains low collision probability among nodes. It allows simultaneous transmissions of a pair of CTS frames or a pair of ACK frames, so that the control overhead remains low. The performances of E2E-KIC MAC have been evaluated analytically, which are confirmed by simulations. The simulation results also show that E2E-KIC (together with its MAC protocol) increases the throughput and reduces the delay, compared to conventional transmission methods that either do or do not support KIC. More detailed signal-level analysis and optimization (e.g., power control and rate adaptation) can be studied in the future.
\section*{Acknowledgments}
This work was supported in part by the National Natural Science Foundation of China (61471109, 9143811, 61302072), the Fundamental Research Funds for the Central Universities (N140405005, N140405007), Liaoning BaiQianWan Talents Program, National High-Level Personnel Special Support Program for Youth Top-Notch Talent.
\bibliographystyle{IEEEtran}
\bibliography{IEEEabrv,SE2EKICMAC}
%\end{spacing}
\end{document}